\documentclass[prd,aps,a4paper,nofootinbib,showpacs,showkeys,twocolumn]{revtex4}  

\newif\ifusesec
\usesectrue  
   
\usepackage{graphicx} 
\usepackage{mathrsfs}
\usepackage{amsmath,amsfonts,amssymb}
\usepackage{multirow}

\newcommand{\beq}{\begin{equation}}
\newcommand{\eeq}{\end{equation}}
\def\rightcontract{\mathop{\hbox{\vrule width0.5pt height6pt
  \vrule height0.5pt width6pt}}}

\begin{document}

\title{Gyroscope precession along unbound equatorial plane orbits around a Kerr black hole}

\author{Donato \surname{Bini}$^1$}
\author{Andrea \surname{Geralico}$^2$}
\author{Robert T. \surname{Jantzen}$^{3}$}

\affiliation{
$^1$Istituto per le Applicazioni del Calcolo ``M. Picone'', CNR, I-00185 Rome, Italy\\
$^2$Astrophysical Observatory of Torino, INAF,
I-10025 Pino Torinese (TO), Italy \\
$^3$Department of Mathematics and Statistics, Villanova University, Villanova, PA 19085, USA
}

\date{\today}

\begin{abstract}
The precession of a test gyroscope along unbound equatorial plane geodesic orbits around a Kerr black hole is analyzed with respect to a static reference frame whose axes point towards the ``fixed stars."
The accumulated precession angle after a complete scattering process is evaluated and compared with the corresponding change in the orbital angle.
Limiting results for the non-rotating Schwarzschild black hole case are also discussed. 
\end{abstract}

\maketitle

\section{Introduction}

In contrast to Newtonian gravitational theory where a test gyroscope orbiting in a stationary gravitational field keeps its spin vector pointing along a fixed direction with respect to the ``fixed stars", in general relativity this direction rotates or ``precesses" due to motion with respect to the stationary field as well as to the Lense-Thirring frame dragging effects of the stationary field itself \cite{thirring1,thirring2,lense-thirring}. Leonard Schiff \cite{schiff} initiated interest in measuring this precession which led to the Gravity Probe B experiment which took many decades to conclude its mission. It confirmed the predictions of general relativity a half century later
\cite{gpb1,gpb2}.

Other than in the context of circular orbits 
\cite{RindlerPerlick:1990,Iyer:1993qa,Bini:1997eb,Bini:2002mh},
very little has been written on the topic of gyroscope precession along more general orbits in curved spacetime \cite{Bini:1997ea,Jantzen:1992rg,Bini:1994}. 
In flat spacetime this effect is intimately connected with Thomas precession 
\cite{Thomas:1926dy,Thomas:1927yu}
which in turn is a consequence of a time-varying direction for a sequence of successive Lorentz boosts along a timelike world line,  coupled with the way in which Lorentz boosts fit into the group of Lorentz transformations
\cite{furry,shelupsky,fischer,goedecke,Ferraro:1999eu,Bakke:2015yia,O'Donnell:2011am,Rhodes:2003id,Han:1987gj}.
At the Lie algebra level, the commutator of any two boost generators along distinct spatial directions leads to a rotation in the plane spanned by the two independent directions. At the Lie group level, two successive boosts are equivalent to a single boost followed by or preceded by a rotation, called a Wigner rotation \cite{Wigner:1939cj}. 

For a time varying boost $B(\lambda)$ along a timelike world line (with parametric equations $x^\alpha=x^\alpha(\lambda)$, $\lambda$ being an affine parameter), the natural Lie algebra derivative  
$B(\lambda)^{-1}\, dB(\lambda)$ or $dB(\lambda)\,B(\lambda)^{-1} $ as appropriate, evaluated using some geometrically defined derivative along the world line and projected into the local rest space of the world line, defines an angular velocity for a time-dependent Wigner rotation along that world line, the ``precession" of the direction of the spin vector of the gyroscope.
For a classical electron in a circular orbit, this is precisely the Thomas precession which ignores the slight deviation of the observer-measured spin vector in both direction and magnitude to maintain its orthogonality to the 4-velocity of its world line.
To make this vague mathematical discourse concrete one has to specify how this boost enters naturally into the problem, and what is its interpretation in terms of observable quantities.

The spin of a test gyroscope in a curved spacetime is well known to undergo Fermi-Walker transport along its world line, which reduces to parallel transport in the case of geodesic motion. Gyroscope precession, however, is a concept which requires a comparison, namely with some preferred reference frame with respect to which the spin vector rotates or ``precesses." In a generic time-varying spacetime with no symmetries, it is hard to imagine how this comparison can be made in a meaningful way.

The easiest case to unambiguously define this mathematical machinery is for the case of planar orbits in an asymptotically flat stationary axially symmetric spacetime, but more general orbits can also be handled. 
For planar motion the spin precession angular velocity vector orthogonal to the orbital plane naturally reduces to a scalar angular velocity precession frequency, another big simplification.
One might as well use the Kerr family of rotating black hole spacetimes including its nonrotating Schwarzschild limit to explicitly evaluate the precession. 
This has been done recently for quasi-elliptical orbits in the equatorial plane of a Kerr spacetime \cite{Bini:2016iym}, namely bound orbits which do not fall into the black hole, but merely precess around the hole. We extend this analysis here to unbound orbits which do not fall into the black hole, namely ``quasi-parabolic" or ``quasi-hyperbolic" orbits which also precess around the hole as they come in from spatial infinity and then go back out in a scattering process. One can then also evaluate the total spin precession at spatial infinity for the entire orbit.

This situation is physically interesting since it 
contains the main characteristics of the more realistic  black-hole black-hole scattering which, when the holes are spinning, may give rise to observational effects associated with the rotation of their spin vectors during the process. 
To get some insight into this more complex problem,
one may consider a test  gyroscope orbiting in the gravitational field of a spinning black hole as a first approximation. 
Finally, the present investigation is also interesting in the context of
recent progress being made in gravitational self-force /gravitational self-torque computations for a particle/gyroscope moving in a perturbed Kerr spacetime \cite{Gair:2005is,Damour:2014afa,Ruangsri:2015cvg}. 
Indeed, in the near future may be possible  to ``analytically"  compute  the full gravitational wave rate of emission by a particle being scattered by the hole, the main difficulty being the existence of a continuous spectrum of frequencies instead of a single frequency as in the simpler case of circular motion. Approaching this challenge will be possible only after the ``background" (Kerr) problem is completely solved and discussed in all its subtle aspects.


\begin{figure}  
\centering
\includegraphics[scale=1.2]{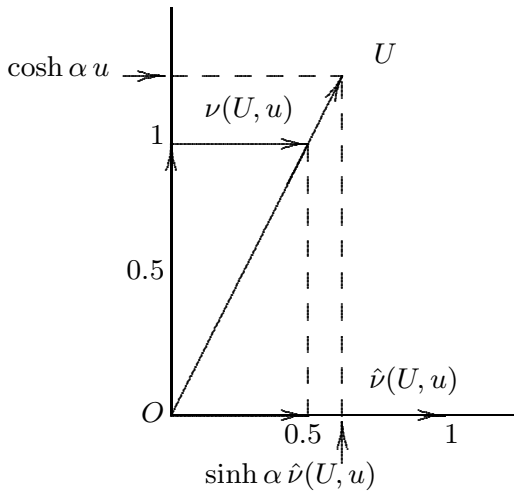} 
\caption{\label{figrelativevelocity}
The relative observer boost plane of the two 4-velocities $u$ and $U$ within the tangent space showing the relative velocity and rapidity hyperbolic decomposition of the projections parallel and perpendicular to $u$. The relative speed is $||\nu(U,u)||=\tanh\alpha$, illustrated here with value 0.5, while the gamma factor is $\gamma(U,u)=\cosh\alpha$.
}
\end{figure}


\begin{figure}  
\centering
\includegraphics[scale=.9]{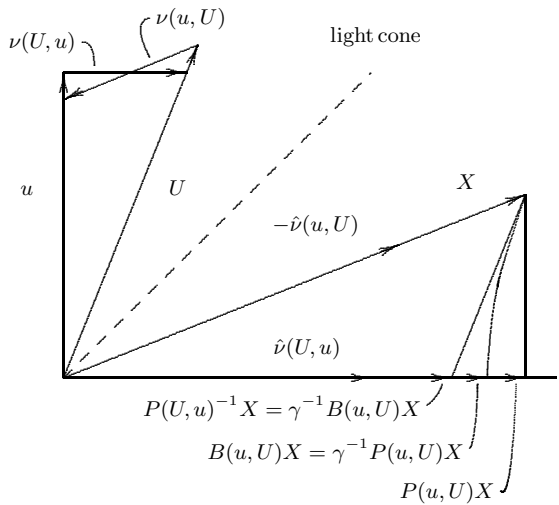} 
\caption{\label{fig:relobsplane}
The relative observer plane, the relative velocities and the associated
relative observer maps between the local rest space directions in that plane, 
for a vector $X\in LRS_U$.
}
\end{figure}

\section{Precession in an asymptotically flat stationary spacetime}

In an asymptotically flat stationary spacetime  consider a ``static" orthonormal frame containing the 4-velocity unit vector field $u=\ell^{-1}\xi$ ($\ell$ being a normalization factor) of a family of world lines which are integral curves of the timelike Killing vector $\xi$, interpreted as the trajectories of ``static test observers", which generically are accelerated. 
By static, we mean at rest with respect to observers at spatial infinity.
Imagining the points on the (nonrotating) sphere at spatial infinity as the 
locations of the ``fixed stars", the light rays arriving from these fixed stars at a point in the spacetime are seen as having fixed directions in the local rest space of the static observers with respect to the static axes in that local rest space. In other words the local static axes (which could be thought of as pointing at three fixed guide stars) determine a local celestial sphere which is rigidly connected to the asymptotic celestial sphere through the stationary symmetry.
If the static observers are carrying test gyroscopes undergoing Fermi-Walker transport, one can directly compare the spin vector with the static axes to evaluate its relative rotation. 
The precession of the actual spin vector with respect to these local axes has a direct interpretation as its precession ``with respect to the fixed stars."

When we consider a timelike world line of a test gyroscope in relative motion with respect to the congruence of static observers, the aberration of the light rays arriving from the celestial sphere at spatial infinity distorts the pattern of fixed stars which are also seen to rotate with respect to a set of axes which are Fermi-Walker transported along the test gyroscope world line. 
 One can remove the effect of the aberration (not literally but suggestively) by actively boosting the static axes to the local rest space of the gyro world line, leaving only the rotation of the fixed stars with respect to the gyro-fixed axes.  This can be interpreted as the precession of the gyroscope spin vector with respect to the fixed stars. Since the boost is an isometry, this is equivalent to boosting the spin vector back to the static observer local rest space and comparing it directly to the static frame (approaching the gyroscope precession problem in this original way was first discussed in \cite{Jantzen:1992rg}, including a number of interesting geometrical details).

The gyros carried along by the static observers precess with the gravitomagnetic precession due to the vorticity of static observer congruence. The additional precession due to the relative motion along general world lines is the total precession. This latter contribution consists of two terms, one due to the 4-acceleration of the world line, and one which remains even for geodesic motion, due to the relative gravitational force associated with the static observers, itself containing both a gravitoelectric and a gravitomagnetic term.

Let $LRS_u$ be the local rest space of the observer $u$ (subspace of the tangent space orthogonal to $u$) and let $U$ be the 4-velocity of the gyro world line in relative motion.
To discuss the boost from $LRS_U$ to $LRS_u$ or viceversa, one needs the machinery of the relative observer boost plane \cite{Jantzen:1992rg}, illustrated in Figs.~\ref{figrelativevelocity} and \ref{fig:relobsplane} (see also \cite{Bini:1997ea,Bini:1997eb} for details). Representing $U$ in terms of its gamma factor, relative velocity and unit direction, and rapidity $\alpha$ with respect to $u$, one has
\begin{eqnarray}
  U &=&\gamma(U,u) \left[ u + \nu(U,u) \right]
\nonumber\\
&=&\gamma(U,u) u +p(U,u)
\nonumber\\
   &=& \cosh(\alpha) u + \sinh(\alpha)\, \hat\nu(U,u)\nonumber\\
&\equiv& B(U,u)\, u
\,,
\end{eqnarray}
where $p(U,u)=\gamma(U,u)\nu(U,u)$ denotes the 4-momentum per unit mass (``of $U$ as seen by $u$'') with
\begin{eqnarray}
\gamma(U,u)&=&(1-||\nu(U,u)||^2)^{-1/2}\,,
\nonumber\\
||\nu(U,u)||&=&(\nu(U,u)\cdot \nu(U,u))^{1/2}\,.
\end{eqnarray}
The boost $B(U,u)$ (``from $u$ to $U$,'' right to left, acting on the left) takes  $u$ to $U$ in their plane, while acting as the identity in the orthogonal complement of this plane.  

Let $P(u) = Id + u\otimes u^\flat$ be the (mixed) projection tensor into $LRS_u$ and identify linear maps from the tangent space into itself with right contraction of the corresponding $1\choose1$-tensor:
$P(u) X =P(u)\rightcontract X$.
We can further decompose the projection into an orthogonal decomposition with respect to the relative velocity direction
\begin{eqnarray}
  P{}^{||}(u,U) &=& \hat\nu(U,u)\otimes \hat\nu(U,u)^\flat\,,
\nonumber\\
  P{}^\perp(u,U) &=& P(u)-\hat\nu(U,u)\otimes \hat\nu(U,u)^\flat\,.
\end{eqnarray}
Let  $B(u,U){}_{\rm(lrs)}=P(u) B(u,U) P(U)$  
be the projection of the inverse boost from $LRS_U$ to $LRS_u$. From Fig.~\ref{fig:relobsplane}, it is clear that this boost merely contracts the parallel direction by a factor of the gamma factor relative to the projection\footnote{A review of the geometrical properties associated with combined boost and projection operations can be found in Ref.~\cite{ferrbini}.} so
\begin{eqnarray}
&& B{}_{\rm(lrs)}(u,U) = P{}^\perp(u,U)+ \gamma(U,u)^{-1} P{}^{||}(u,U)
\\
&&\qquad  =P(u)-(1-\gamma(U,u)^{-1})\hat\nu(U,u)\otimes \hat\nu(U,u)^\flat
\,.\nonumber
\end{eqnarray}
This relative boost contains all the information about the spin precession of a test gyroscope moving along its orbit, and its appropriate Lie algebra derivative along the world line determines the angular velocity of the rotation of the direction of the spin vector relative to the static frame of the static observers.

\section{Equatorial plane orbits around a Kerr black hole}

Consider the Kerr metric written in standard Boyer-Lindquist coordinates $(x^\alpha)=(t,r,\theta,\phi)$
\cite{Misner:1974qy,Chandrasekhar:1985kt}
\begin{eqnarray}\label{K1}
d s^2 &=& g_{\alpha\beta}dx^\alpha dx^\beta\nonumber\\
&=&-d t^2+\frac{\Sigma}{\Delta} \,d r^2+\Sigma \,d\theta^2 +(r^2+a^2)\sin^2\theta \,d\phi^2\nonumber\\
&&
+\frac{2Mr}{\Sigma}(dt-a\sin^2\theta \,d\phi)^2\,,
\end{eqnarray}
where $a=J/M$ is the specific angular momentum of the source (with $\hat a=a/M$ dimensionless)  and
\beq\label{K2}
\Sigma=r^2+a^2\cos^2\theta\,,\qquad \Delta=r^2-2Mr+a^2\,.
\eeq
The inner and outer horizon radii are at $r_\pm=M\pm\sqrt{M^2-a^2}$.
Units are chosen here such that $G=c=1$. The static observers move along the time coordinate lines with 4-velocity $u\equiv(-g_{tt})^{-1/2}\, \partial_t$ aligned with the Killing vector field $\partial_t$ and play a fundamental role in the spin precession as seen by observers at rest with respect to the coordinates far from the black hole. 

The Boyer-Lindquist spatial coordinates $r,\theta,\phi$ are spherical-like coordinates whose normalized coordinate spatial frame can be boosted to the local rest space of the static observer 4-velocity $u$ along the azimuthal direction to yield a spherical triad $E_i,i=1,2,3$ defining the local celestial sky tied to the asymptotically flat (nonrotating) celestial sphere by the stationary symmetry. At the equatorial plane $\theta=\pi/2$, this frame is explicitly \cite{Jantzen:1992rg,bcjkorea}
\begin{eqnarray} 
\label{frame_thd}
u &=& \frac{1}{N}\,\partial_t \,,\quad 
E_1 =\sqrt{\frac{\Delta}{\Sigma}} \,\partial_r\,,\nonumber\\ 
E_2&=& \frac{1}{\sqrt{\Sigma}} \,\partial_\theta\,\quad
E_3= \frac{N}{\sqrt{\Delta}} \,\left[\partial_\phi+\frac{2aM}{rN^2} \,\partial_t\right] \,,
\end{eqnarray}
where $N=\sqrt{1-2M/r}$.
The orthonormal spatial frame vectors $E_1,E_2,E_3$ satisfy the usual cross-product algebra with respect to the $\times_u$ cross product operation on $LRS_u$, with $E_2$ pointing ``downward" at the equatorial plane.
One might even introduce a ``Cartesian frame" within the equatorial plane with $E_z=-E_2$ and rotating the remaining two spherical frame vectors by the angle $\phi$ in the clockwise direction in that plane
\beq\label{cartesianframe}
  \begin{pmatrix} E_x & E_y\end{pmatrix}
  = \begin{pmatrix} E_1 & E_3\end{pmatrix} 
    \begin{pmatrix} \cos\phi & \sin\phi\\ -\sin\phi & \cos\phi \end{pmatrix}\,.
\eeq
This provides a reference frame to compare a moving gyroscope spin vector direction and thus define a precession along its orbit, once its relative motion with respect to the static observers is taken into account with a relative boost.

The acceleration $a(u)=\nabla_u u$ and vorticity vector $\vec \omega(u)$ (i.e., the spatial dual of the vorticity tensor, defined as the antisymmetric (ALT) part of the projected covariant derivative of $u$, $\omega(u)={\rm ALT}P(u)\nabla u$) of the static observer congruence at the equatorial plane are
\beq
\label{eq_a_omega}
a(u) = \frac{M\sqrt{\Delta}}{r^3N^2 } \,E_1\,, \quad
\vec\omega(u) =  -\frac{aM}{r^3N^2} \,E_2 \,.
\eeq
These point radially outward (to resist falling into the black hole) and vertically upward at the equatorial plane (corresponding to frame dragging in the positive azimuthal direction when $a>0$), respectively.

Timelike geodesic world lines in this metric  $x^\alpha=x^\alpha (\tau)$ parametrized by the proper time $\tau$ have a 4-velocity $U^\alpha=dx^\alpha/d\tau$ whose coordinate components satisfy
\begin{eqnarray}
\label{geosgen}
\frac{d t}{d \tau}&=& \frac{1}{\Sigma}\left[aB+\frac{(r^2+a^2)}{\Delta}P\right]\,,\nonumber \\
\frac{d r}{d \tau}&=&\epsilon_r \frac{1}{\Sigma}\sqrt{R}\,,\nonumber \\
\frac{d \theta}{d \tau}&=&\epsilon_\theta \frac{1}{\Sigma}\sqrt{\Theta}\,,\nonumber \\
\frac{d \phi}{d \tau}&=& \frac{1}{\Sigma}\left[\frac{B}{\sin^2\theta}+\frac{a}{\Delta}P\right]\,,
\end{eqnarray}
where $\epsilon_r$ and $\epsilon_\theta$ are sign indicators, and
\begin{eqnarray}
\label{geodefs}
P&=& E(r^2+a^2)-La\,,\nonumber\\
B&=& L-aE \sin^2\theta\,, \nonumber\\
R&=& P^2-\Delta (r^2+K)\,,\nonumber\\
\Theta&=&K-a^2\cos^2\theta-\frac{B^2}{\sin^2\theta}\,,
\end{eqnarray}
where $K$ is Carter's constant  associated with the symmetric Killing 2-tensor 
of the Kerr spacetime~\cite{Carter:1968ks}
and $E$ and $L$ are the conserved energy and angular momentum per unit mass
associated with the Killing vector fields $\partial_\phi$ and $\partial_t$ of a test particle in geodesic motion.
Note that $E$ and $L/M$ are dimensionless. 

We are interested here in equatorial plane orbits, i.e., orbits at $\theta=\pi/2$ with $K=(L-a\,E)^2\equiv x^2$ (with $\hat x=x/M$ dimensionless) so that
\begin{eqnarray}\label{eqsmotionequatorial}
\label{eq_t}
\Delta\,r^2 \,\frac{dt}{d\tau} &=& 
(Er^2-ax)(r^2+a^2)+ \Delta a x\,, \nonumber\\
\label{eq_r}
r^4 \,\Big(\frac{dr}{d\tau}\Big)^2  &=& (Er^2  -a x)^2-\Delta(r^2+x^2)\,, \\
\label{eq_phi}
\Delta\,r  \,\frac{d\phi}{d\tau}&=& 
rL-2Mx 
\nonumber\\
&=& r\left[ L \left(1-\frac{2M}{r}\right) +aE \left(\frac{2M}{r}\right)\right]
\,.\nonumber
\end{eqnarray} 
When the orbital angular momentum $L$ and the rotation $a$ of the black hole are of the same/opposite sign (prograde or corotating orbits versus retrograde or counterrotating orbits), the orbital angular velocity $\dot\phi\equiv d\phi/d\tau$ is increased/decreased. 
Eq.~\eqref{eq_phi} implies that ${d\phi}/{d\tau}$ changes its sign when
\beq
\frac{L}{aE}=-\frac{2M}{r-2M}\,,\quad r_s=2M \left(1-a\frac{E}{L}  \right)\,.
\eeq
If ${\rm sgn}(aL)>0$ (prograde),   $r_s$ is always inside the outer horizon, whereas  if ${\rm sgn}(aL)<0$ (retrograde), $r_s$ can lie in the region outside but near the horizon. This is irrelevant if $r_s$ is less than minimum radius along an unbound orbit.

If $a\ge 0$,  then prograde (corotating) and retrograde (counterrotating) orbits correspond respectively to  $ \dot\phi>0$ and $\dot\phi<0$.
Formulas valid for retrograde orbits can be obtained from those for prograde orbits by $a\to -a$ and $L \to - L$, under which $x\to -x$.

At the equatorial plane the downward vertical unit vector $E_2$ along $\theta$ is covariant constant within the plane, and the precession of a test gyroscope in such a planar orbit only undergoes a rotation in the 2-plane of the radial $r$ and azimuthal $\phi$ directions. These static observer frame directions are all locked to the observers at rest at spatial infinity, and so provide a natural way to measure the spin precession as seen by distant observers, modulo the boost between the local rest space of the gyro and that of the static observers tied to the coordinate grid needed to compensate for the relative motion.

With respect to the frame \eqref{frame_thd}, the geodesic 4-velocity can be represented as $U=\gamma(U,u)[u+\nu(U,u)]$, with 
\beq
  \nu(U,u) = \nu(U,u)^1 E_1 + \nu(U,u)^3 E_3 \,,
\eeq 
and, suppressing the $(U,u)$ qualifier for simplicity, 
\beq
\gamma =\frac{E}{N}\,,\quad
\gamma \nu ^1=\frac{r}{\sqrt{\Delta}}\frac{d r}{d \tau}\,,\quad
\gamma \nu ^3=\frac{\sqrt{\Delta}}{N}\frac{d \phi}{d \tau}\,.
\eeq

The radial variable along those orbits which have a minimum radius and nonconstant $\phi$ can be expressed in the form \cite{Chandrasekhar:1985kt}
\beq\label{rversuschi}
r =\frac{Mp}{1+e\cos \chi}\,, 
\eeq
where $\chi$ is a new function of the proper time along world line of the gyro,
$e\geq0$ is the eccentricity parameter, and $Mp$ is the semi-latus rectum. Note that $p$ is dimensionless, as is its reciprocal $u_p=1/p$.  
with. The value $\chi=0$ corresponds to the periastron radius,
\beq
r_{\rm (per)}=\frac{Mp}{1+e}\,.
\eeq

\begin{figure*}
\centering
\includegraphics[scale=0.35]{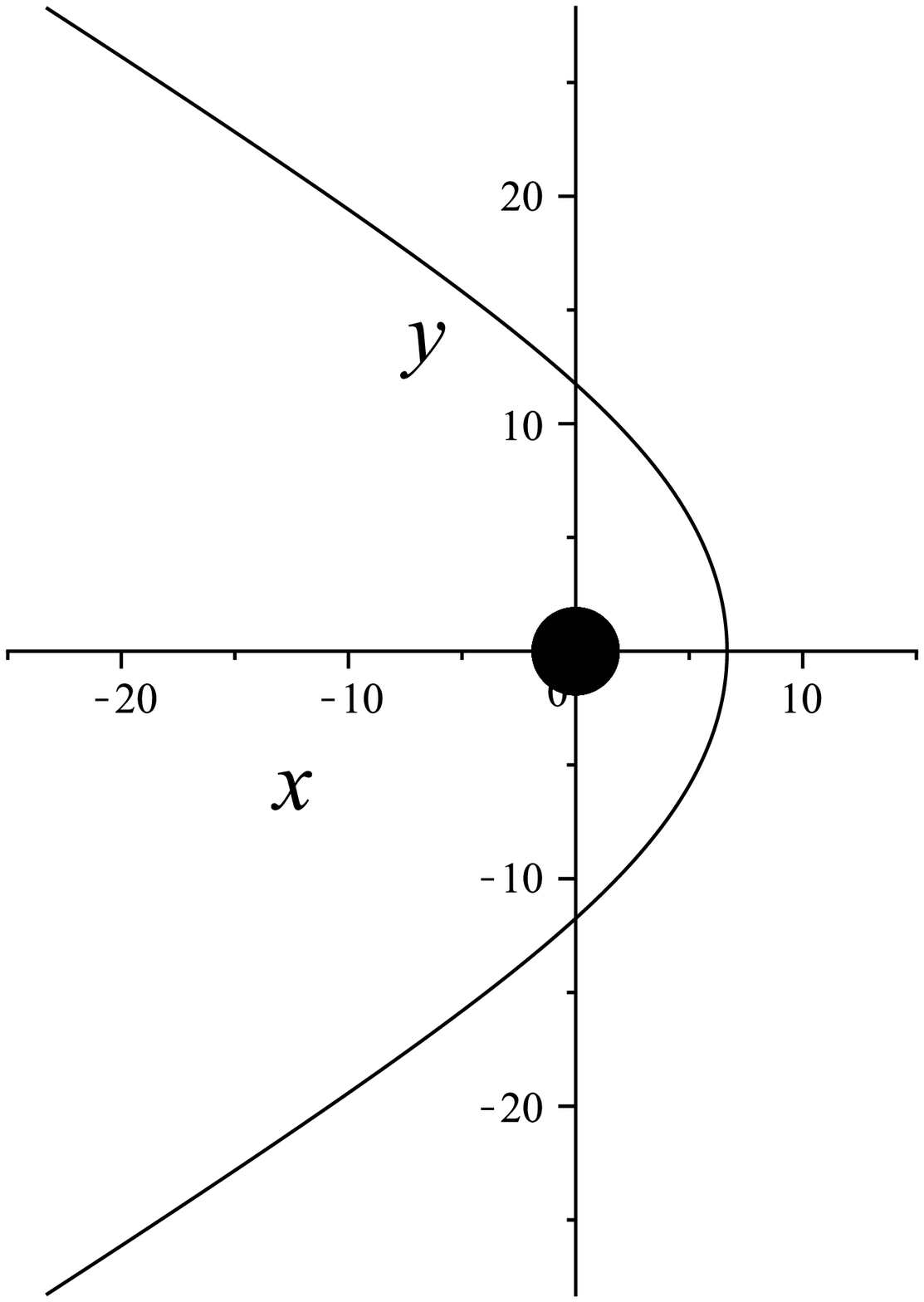}
\includegraphics[scale=0.35]{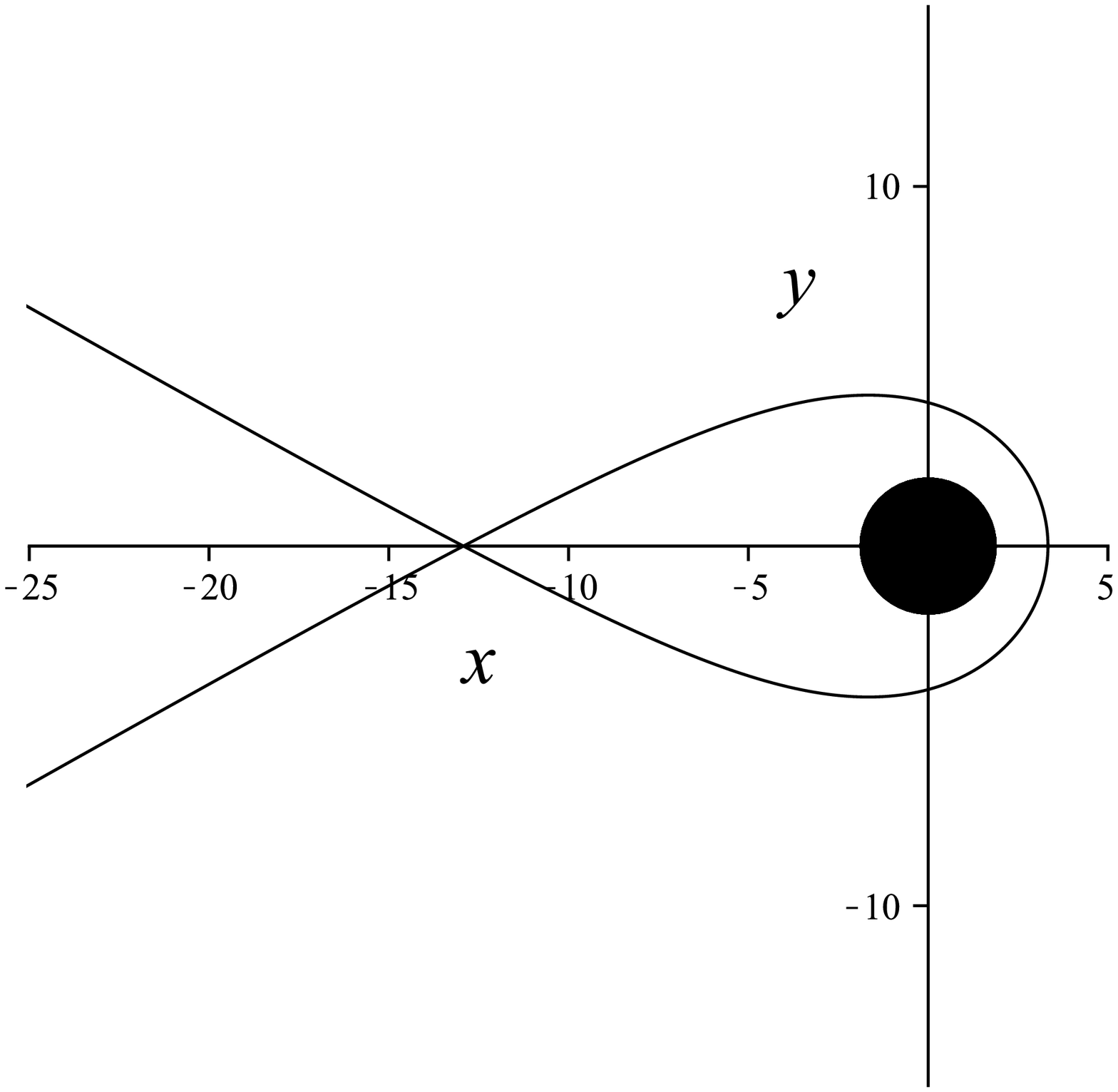}
\hbox to \hsize{\hfil $a)$ \hfil $b)$ \hfil}
\caption{\label{fig_orbit}The geodesic orbit (here $x=(r/M)\cos\phi$ and $y=(r/M)\sin\phi$ are Cartesian-like coordinates) is shown for $L>0$, ${\hat a}=0.5$, $e=2$ (so that $\chi_{\rm(max)}=2\pi/3$) and $u_p=0.05$ (left panel) and $u_p=0.1$ (right panel). In the former case
we have $\hat x \approx 4.7410$, $L/M \approx 5.2831$, $E \approx 1.0841$; in the latter case instead $\hat x \approx4.0384$, $L/M \approx4.6398$, $E \approx1.2028$. The solid circles correspond to the outer horizon.
}\label{orbits}
\end{figure*}

In this paper we are interested in unbound orbits, i.e., with eccentricity $e\ge1$, starting far from the hole at radial infinity, reaching a minimum approach distance $r_{\rm per}$ from the hole (periastron), and then coming back to radial infinity,
corresponding to 
$\chi\in [-\chi_{\rm (max)}, \chi_{\rm (max)}]$, $\chi_{\rm (max)}=\arccos(-1/e)$.
The special case $E=1=e$ corresponds to a precessing parabolic orbit (say, ``parabolic-like"'), while $E>1,e>1$ corresponds to precessing hyperbolic orbits (say, ``hyperbolic-like"').
For simplicity we will assume that $d\phi/d\tau>0$ and allow $-1\le \hat a \le 1$, so the sign of $\hat a$ determines prograde ($+$) and retrograde ($-$) orbits.
Fig.~\ref{orbits} shows two examples of hyperbolic-like orbits with $\chi_{\rm (max)}=2\pi/3$ and  $u_p=0.05,0.1$.


The conserved quantities $E$ and $\hat x$ can be expressed in terms of $p$ and $e$ as follows \cite{Glampedakis:2002ya,Bini:2016iym}
\begin{eqnarray}
 E&=& -\frac{p-3-e^2}{2\hat a   p} \hat x 
-\frac{ (\hat a^2-p)}{2\hat a }\frac{1}{\hat x} \,,\nonumber\\
\hat x{}^2 &=&\frac{- {\mathcal N}-{\rm sgn}(a) \sqrt{{\mathcal N}^2-4CF}}{2F}\,, 
\end{eqnarray}
where the dimensionless coefficients $F$, $N$ and $C$ given by 
\begin{eqnarray}
 F  &=&  \left(1-\frac{ 3+e^2 }{p} \right)^2-\frac{4\hat a^2(1-e^2)^2}{p^3}  \,,\nonumber\\
-\frac{{\mathcal N}}{2}&=&(p-3-e^2)+\hat a^2\left(1 +\frac{1+3e^2}{p}\right)\,, \nonumber\\
C&=&(\hat a^2-p)^2\,.
\end{eqnarray}
[See Ref.~\cite{Bini:2016iym} for a discussion of the correlation between the signs in the roots of the biquadratic for $\hat x$ and the sign of $\hat a$.]

Eq.~\eqref{eq_r} can be rewritten in  factored form
\beq
\label{r_eq2}
\left(\frac{dr}{d\tau}\right)^2=\frac{(E^2-1)}{r^3}(r-r_3)(r-r_{2})(r-r_{\rm per})\,,
\eeq
where 
\beq
\label{eq_ni_r3EX}
\frac{r_2}{M}=\frac{Mp}{1-e}\,,\qquad
\frac{r_3}{M} = \frac{2\hat x{}^2(e^2-1)}{p^2(E^2-1)}\,. 
\eeq
The motion is confined to $r\ge r_{\rm per}$. 
As in the case of bound orbits, when $r_3 = r_{\rm per}$ the effective potential for radial motion has a critical point with a negative second derivative at the periastron corresponding to an unstable circular orbit radius $r_c$, making the eccentric orbit at that energy marginally stable \cite{Glampedakis:2002ya}.
This condition on allowed values of $(e,p)$ determines the \lq\lq separatrix,'' whose parametric equations are given by \cite{Levin:2008yp}
\begin{eqnarray}\label{separatrix}
e^{\rm sep}&=&-\frac{r_c^2-6Mr_c-3a^2\pm8a\sqrt{Mr_c}}{\Delta_c}\,, \nonumber\\
p^{\rm sep}&=&\frac{4r_c}{\Delta_c}(\sqrt{Mr_c}\mp a)^2\,,
\end{eqnarray}
with $\Delta_c=\Delta(r_c)$.
These may be re-expressed in terms of the parameter $u_p=1/p$ using $u_p(1+e)=M/r_c$.

Finally, we list below the $t$ and $\phi$ geodesic equations parametrized by $\chi$, i.e.,

\begin{widetext}

\begin{eqnarray}
\label{dtdchi_dphidchi}
\frac{dt}{d\chi}&=&\frac{M}{u_p^{3/2}}\frac{E +E\hat a^2 u_p^2 (1+e\cos\chi)^2   -2 \hat a u_p^3\hat x (1+e\cos\chi)^3}{(1+e\cos \chi )^2[1+u_p^2 \,\hat x{}^2 (e^2-2 e\cos \chi -3) ]^{1/2}
[1-2 u_p(1+ e\cos \chi) +a^2 u_p^2(1+ e\cos \chi)^2  ]}
\,,\nonumber\\
\frac{d\phi}{d\chi}&=& u_p^{1/2}\frac{ \hat x + \hat a E - 2 u_p \hat x (1+  e\cos \chi) }{[1+u_p^2 \,\hat x{}^2 (e^2-2 e\cos \chi -3) ]^{1/2}
[1-2 u_p(1+ e\cos \chi) +a^2 u_p^2(1+ e\cos \chi)^2  ]}
\,,
\end{eqnarray}
with
\begin{eqnarray}
M \frac{d\chi}{d\tau} &=&  u_p^{3/2}(1+e\cos \chi )^2
[1+u_p^2\, \hat x{}^2 ( e^2-2 e\cos\chi-3)]^{1/2}\,.
\end{eqnarray}

\end{widetext}

\begin{figure*}
\centering
\includegraphics[scale=0.35]{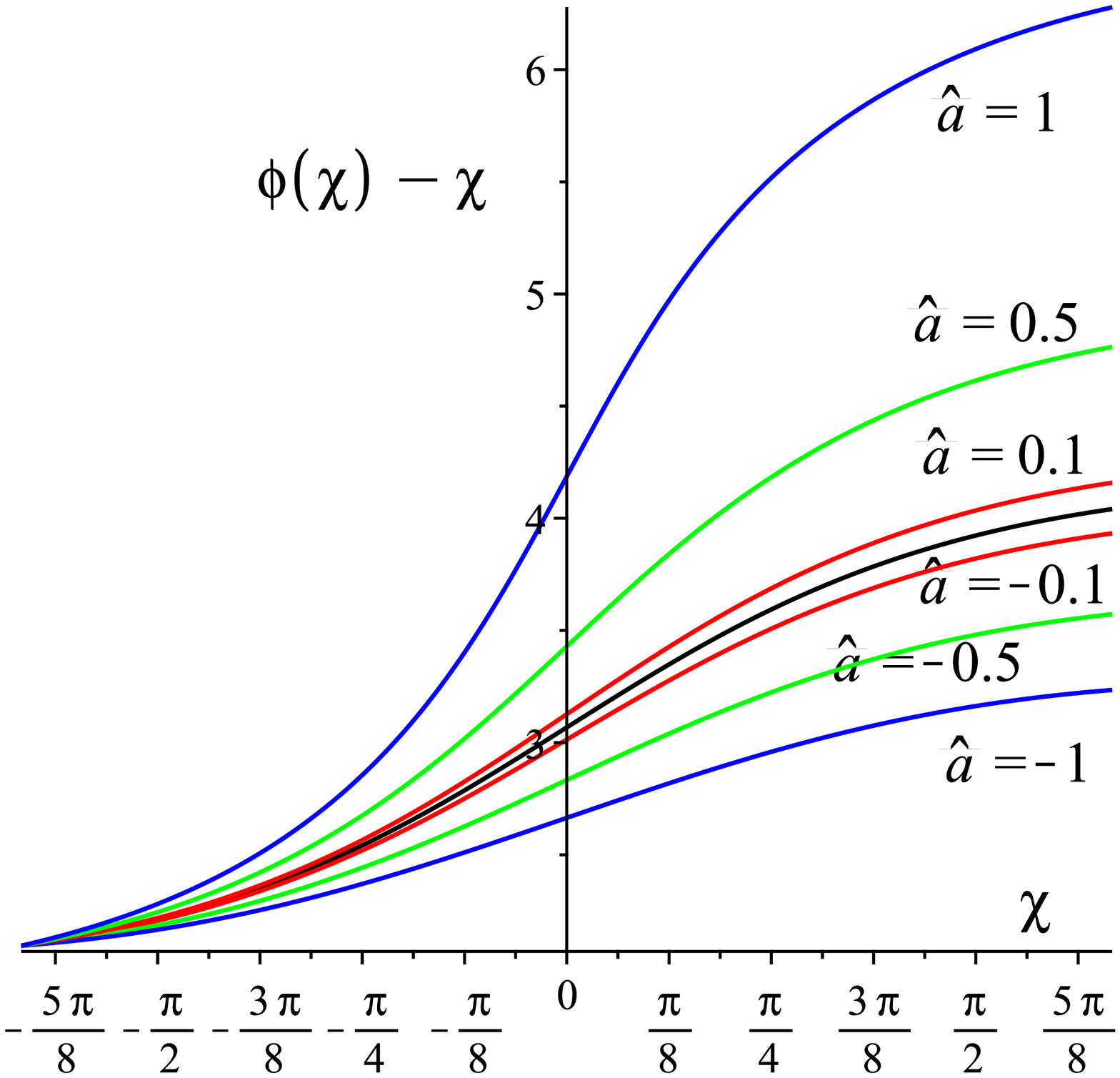}\qquad
\includegraphics[scale=0.35]{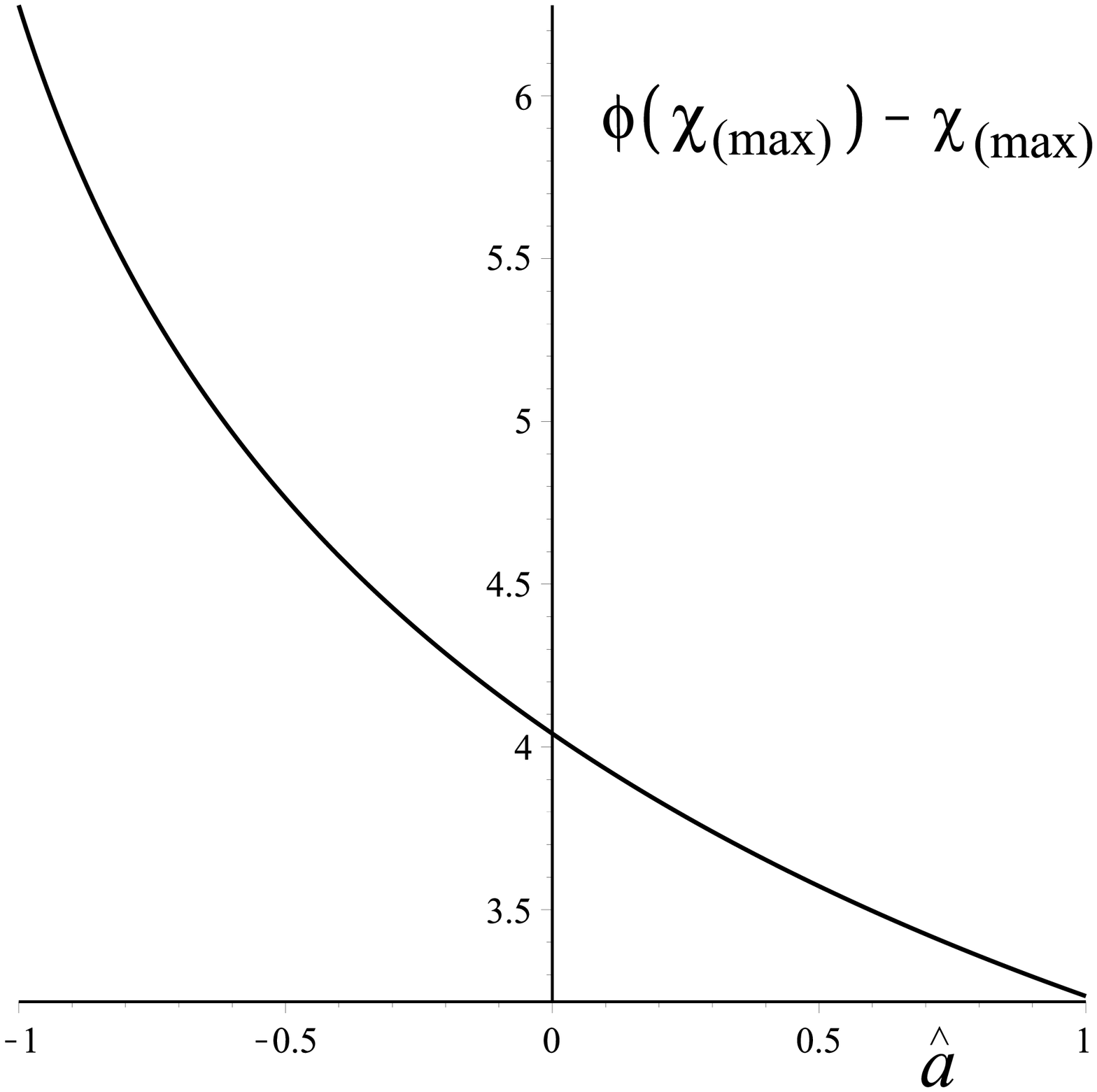}
\caption{\label{fig_bob}
Left panel.
The quantity  $\phi(\chi)-\chi$ is plotted in degrees as a function of $\chi$ for $e=2$ and $u_p=1/15$ and the selected values $\hat a=[0,\pm 0.1,\pm 0.5,\pm 1]$.
Right panel.
The quantity  $\phi(\chi_{\rm(max)})-\chi_{\rm(max)}$ is plotted as a function of $\hat a$ for $e=2$ and $u_p=1/15$.
}
\end{figure*}

Unbound geodesic orbits in the equatorial plane which are not captured by the black hole
start at infinite radius at the azimuthal angle $\phi=\phi(-\chi_{\rm (max)})$, 
the radius decreases to its periastron value at $\phi=\phi(0)$ and then returns in a symmetrical way back to infinite value at $\phi=\phi(\chi_{\rm (max)})$, undergoing a total increment of $\Delta\phi =\phi(\chi_{\rm (max)})-\phi(-\chi_{\rm (max)})$. If we choose $\phi(0)=0$,
then $\Delta\phi =2\phi(\chi_{\rm (max)})$, and the deflection angle from the original direction of the orbit is the complementary angle $\pi-2\phi(\chi_{\rm (max)})$.
The difference $\Delta\phi - 2\chi_{\rm (max)}=2(\phi(\chi_{\rm (max)})-\chi_{\rm (max)})$ is the additional precession of the hyperbolic-like orbit relative to the corresponding Newtonian hyperbolic orbit with the same parameters $(e,p)$.
Fig.~\ref{fig_bob} shows the behavior of this quantity for selected values of these parameters.


\subsection{Explicit solution of orbital equations}

One can solve the orbital equation $\phi=\phi(r)$ as follows starting from
the equations for $r$ and $\phi$ in terms of the dimensionless inverse radial variable $u=M/r$, i.e.,
\begin{eqnarray}
\left(\frac{du}{d\tau}\right)^2&=&\frac{2\hat x^2}{M^2}u^4(u-u_1)(u-u_2)(u-u_3)\,,\nonumber\\
\frac{d\phi}{d\tau}&=&\frac{2\hat x}{M\hat a^2}u^2\frac{u_4-u}{(u-u_+)(u-u_-)}\,,
\end{eqnarray}
where
\beq
u_1=-u_p(e-1)\,,\quad
u_2=u_p(e+1)\,,\quad
u_3=\frac{M}{r_3}\,,
\eeq
and
\beq
u_\pm=\frac{M}{r_\pm}\,,\quad
u_4=\frac{2L}{x}\,,
\eeq
which can be combined to yield
\beq
\label{phidiusol}
\frac{du}{d\phi}=\frac{\hat a^2}{\sqrt{2}}\frac{(u-u_+)(u-u_-)}{u_4-u}\sqrt{(u-u_1)(u-u_2)(u-u_3)}\,.
\eeq
For hyperbolic orbits we have $u_1<0<u\leq u_2<u_3$, so that integration gives
\beq
\phi(u)=\frac{\sqrt{2}}{\hat a^2}\int_{u_2}^uf(z)\frac{dz}{\sqrt{(z-u_1)(u_2-z)(u_3-z)}}\,,
\eeq
with $\phi(u_2)=0$ (periastron) and
\beq
f(z)=\frac{u_4-z}{(z-u_+)(z-u_-)}\,.
\eeq
Using the decomposition
\beq
f(z)=\frac1{u_+-u_-}\left(\frac{u_4-u_+}{z-u_+} 
-\frac{u_4-u_-}{z-u_-}\right)\,,
\eeq
Eq. \eqref{phidiusol} can be expressed in terms of elliptic integrals (defined below) through formula $3.137(4)$ on 
p.~258 of \cite{Gradshteyn} as 

\begin{widetext}

\beq
\label{phidiusol2}
\phi(u)=\frac{2\sqrt{2}}{\hat a^2(u_+-u_-)\sqrt{u_3-u_1}}\left[
\frac{u_4-u_+}{u_1-u_+}\left(\Pi(\alpha,\beta_+,m)-\Pi(\beta_+,m)\right)
-\frac{u_4-u_-}{u_1-u_-}\left(\Pi(\alpha,\beta_-,m)-\Pi(\beta_-,m)\right)
\right]\,,
\eeq
where
\beq
m=\sqrt{\frac{u_2-u_1}{u_3-u_1}}\,,\quad
\alpha=\sqrt{\frac{u-u_1}{u_2-u_1}}\,,\quad
\beta_\pm=\frac{u_2-u_1}{u_\pm-u_1}\,.
\eeq
The total change in $\phi$ for the complete scattering process is then given by $2\phi(0)$ determined by Eq.~\eqref{phidiusol2} with $\alpha=\alpha(0)=\sqrt{-u_1/(u_2-u_1)}$.

In the Schwarzschild limit we have
\beq
m=2 \sqrt{\frac{eu_p}{1-6u_p +2eu_p}}\,,\quad
\alpha=\sqrt{\frac{u}{2eu_p}+\frac{e-1}{2e}}\,,\quad
\beta_+=\frac{4eu_p}{1-2u_p +2eu_p}\,,\quad
\beta_-=0\,,
\eeq

\end{widetext}
and $\Pi(\alpha,0,m)=F(\alpha,m)$ and $\Pi(0,m)=K(m)$, so that 
\beq
\label{phiuschw}
\phi(u)=\frac{m}{\sqrt{eu_p}}\left[ K(m)-F(\alpha,m) \right]\,,
\eeq
and hence
\beq
\label{phi0schw}
\phi(0)=\frac{m}{\sqrt{eu_p}}\left[ K(m)-F\left(\sqrt{\frac{e-1}{2e}},m  \right) \right]\,.
\eeq
Note that the first term in Eq. \eqref{phidiusol2} vanishes in this limit, since $u_4=1/2=u_+$.

Here $K(k)$ and $F(\varphi,k)$ are the incomplete and complete elliptic integrals of the first kind, respectively, defined by
\beq
F(\varphi,k)=\int_0^{\varphi}\frac{dz}{\sqrt{1-k^2\sin^2z}}\,,\quad
K(k)=F(\pi/2,k)\,,
\eeq 
whereas
\begin{eqnarray}
\Pi(\phi,n,k)&=&\int_0^{\varphi}\frac{dz}{(1-n\sin^2z)\sqrt{1-k^2\sin^2z}}\,,\nonumber\\
\Pi(n,k)&=&\Pi(\pi/2,n,k)\,,
\end{eqnarray}
are the incomplete and complete elliptic integrals of the third kind, respectively~\cite{Gradshteyn}.

\section{Marck's parallel propagated frame rotated from a preliminary Frenet-Serret frame}

Marck constructed an orthonormal frame containing the gyro 4-velocity $U=e_0$ which is parallel transported along an arbitrary geodesic in the Kerr spacetime \cite{marck1,marck2} using Kerr's Killing-Yano tensor 2-form $f$.   
In the equatorial plane the second frame vector is then obtained by forming the unit spacelike 1-form 
\begin{equation}\label{K29}
e_\mu{}_{2}=\frac{1}{x}f_{\mu\nu} U^\nu =  r \delta^\theta{}_\mu
\end{equation}
which is orthogonal to $e_{0}$  and is parallel propagated along the geodesic orbit.  
Marck then completed these first two vector fields to an orthonormal frame adapted to $U$ by adding the two vector fields whose corresponding 1-forms (indicated by $\flat$) at the equatorial plane are \cite{Bini:2011zzc,Bini:2016xqg} 
\begin{eqnarray}
   \tilde e_{1} {}^\flat
&=& \frac{x}{\sqrt{x^2+r^2}} \left[ -r \dot r \, (dt-a\,d\phi)+\frac{r}{\Delta} (r^2 E-a x)\,dr\right] \,,\nonumber\\
   \tilde e_{3} {}^\flat
 &=& \frac{x}{\sqrt{x^2+r^2}} \left[
\, \frac{r^2 }{\Delta}\,\dot r dr  -
\, \frac{(r^2 E-a x)}{r^2} (dt -a d\phi) \right]\nonumber\\
&&
- \frac{\sqrt{x^2+r^2}}{r^2} 
\,  [a\, dt -(r^2+a^2)\, d\phi] \,,
\end{eqnarray}
where $\dot r = dr/d\tau$ is given by \eqref{eq_r}.
The associated vectors instead read
\begin{eqnarray}
\tilde e_1 &=& \frac{1}{\Delta \sqrt{r^2+x^2}}\left[ r\dot r (r^2+a^2)\partial_t +\frac{(Er^2+ax)\Delta}{r}\partial_r\right. \nonumber\\
&& \left.  +ar \dot r \partial_\phi \right]\nonumber\\
\tilde e_3 &=& \frac{1}{\Delta \sqrt{r^2+x^2}} \left[\frac{x(Er(r^2+a^2)-2Max)+ar\Delta}{r} \partial_t \right. \nonumber\\
&& \left.+x\Delta\dot r \partial_r+\frac{x(raE+(r-2M)x)+r\Delta}{r}\partial_\phi \right]\,.
\end{eqnarray}

This initial non-parallel propagated  frame\\ $\{e_0,\tilde e_1,e_2,\tilde e_3\}$ is closely related (through two successive relative velocity boosts \cite{Bini:2016iym}) to the Carter orthonormal frame 
\cite{Carter:1968ks} which is the key to the geometrical properties of the Kerr spacetime, unlocking the geodesic equations through their separability, as well as Maxwell's equations and other linear spin field equations.  The timelike and radial Carter frame vectors span the 2-plane containing the principal null directions of the spacetime, while its temporal and azimuthal frame vectors span the orthogonal 2-plane containing the  Killing vectors~\cite{ferrbini}. This frame aligns the electric and magnetic parts of the Killing 2-form and diagonalizes the electric and magnetic parts of the Riemann curvature tensor (equal to the Weyl tensor), while the magnetic part vanishes on the equatorial plane. The Marck frame in turn is the key to parallel transport along geodesics.

For Marck's preliminary frame the electric and magnetic parts of the curvature (the sign of the magnetic part of the Riemann tensor depends on the sign convention chosen for the unit volume $4$-form used to define the duality  $*$ operation)
\begin{eqnarray}
E(U)_{\alpha\gamma}=R_{\alpha\beta\gamma\delta}   U^\beta    U^\delta\,,  \nonumber\\
H(U)_{\alpha\gamma}={}^*\kern-1pt R_{\alpha\beta\gamma\delta}   U^\beta   U^\delta \,, 
\end{eqnarray}
evaluated along the geodesic world line 4-velocity $U$, have the following simple expressions 
\begin{eqnarray}
E(U)&=&\frac{M}{r^5}\left[ -(3x^2 +2r^2)\tilde e_1 \otimes \tilde e_1 +(3x^2 +r^2)\tilde e_2 \otimes \tilde e_2\right.\nonumber\\
&&\left.  +r^2 \tilde e_3\otimes \tilde e_3 \right]\,,\nonumber\\
H(U)&=& -\frac{3M}{r^5}x \sqrt{r^2+x^2}  (\tilde e_1 \otimes \tilde e_2+\tilde e_2 \otimes \tilde e_1)\,.
\end{eqnarray}
The electric part of Riemann is still diagonal here, while the magnetic part has a canonical off-diagonal form.
This diagonalization was mentioned in Ref.~\cite{Akcay:2016dku} in their discussion of  bound equatorial plane geodesic orbits. This property seems to be the main geometrical characterization of Marck's frame for general orbits.

This preliminary Marck frame is a degenerate  Frenet-Serret frame along the geodesic
(but in the order $\{e_0,\tilde e_1,\tilde e_3,e_2\}$ with $\kappa=0,\tau_1=\mathcal{T},\tau_2=0$ in Ref.~\cite{Bini1999})
\begin{eqnarray}
\frac{De_{0}}{d\tau} &=& 0\,,\qquad \frac{De_{2}}{d\tau} = 0\,,\nonumber\\
\frac{D\tilde e_{1}}{d\tau} &=&{\mathcal T} \tilde e_{3}  \,,\qquad 
\frac{D\tilde e_{3}}{d\tau} = -{\mathcal T} \tilde e_{1}\,,
\end{eqnarray}
with Frenet-Serret angular velocity vector $\omega_{\rm(FS)}=-{\mathcal T} e_{2}$, so that
\beq
\label{eq_FS_U}
  \frac{D\tilde e_{i}}{d\tau} = \omega_{\rm(FS)}\times_U\tilde e_{i}\,.
\eeq
By direct evaluation one finds the angular velocity ${\mathcal T}$ of the gyro-fixed axes with respect to the preliminary Marck axes (in the clockwise direction when $\mathcal{T}>0$) is \cite{marck1,marck2}
\beq\label{K31bis}
 {\mathcal T} = \frac{d\Psi}{d\tau}
=\frac{a+E x}{r^2+x^2}
= \frac{a(1-E^2)+EL}{r^2+(L-aE)^2}
\,.
\eeq

When $\mathcal{T}>0$, the Marck frame vectors $\tilde e_1,\tilde e_3$ rotate in the counterclockwise azimuthal direction with respect to parallel transported axes.
Rotating them by a clockwise rotation angle $\Psi$ in this plane
to get Marck's final parallel propagated frame $\{e_\alpha\}$,
\beq\label{K30}
\left(\begin{array}{c} e_1 \\ e_3 \end{array}\right)
= R(\Psi) \left(\begin{array}{c} \tilde e_1 \\ \tilde e_3 \end{array}\right)
\equiv\left(\begin{array}{cc}  \cos \Psi &- \sin \Psi \\ \sin \Psi & \cos \Psi\end{array}\right)
\left(\begin{array}{c} \tilde e_1 \\ \tilde e_3 \end{array}\right)
\,,
\eeq
i.e., the tilde frame is rotated by the angle $\Psi$ in the counterclockwise direction with respect to the parallel propagated frame in that plane. A parallel transported spin vector has constant components in this frame.

For a circular orbit at constant $r$, $\mathcal{T}$ is a constant frequency leading to a uniform rotation of the spin vector.
For unbound orbits with $E>1$, the two terms in the final right hand side of \eqref{K31bis} show that the frame dragging effect associated with $a$ reverses direction with respect to the bound orbits where $E<1$ but apart from sign is aligned with the rotation of the black hole independent of the azimuthal orbital direction, while the orbital precession term associated with $L$ is always in the same direction as that orbital direction.
For $E=1$ orbits the frame dragging term does not directly contribute, and when in addition $L=0$, the total precession term $\mathcal{T}$ vanishes, which means that the spin direction is locked onto the Marck frame for purely radial infall in the Schwarzschild case.
 
The Marck frame vector $\tilde e_{1}$ itself is locked to the radial direction $E_1=1/\sqrt{g_{rr}}\,\partial_r$  in the spherical grid of the static observers following the time lines, 
differing only but a boost due to the radial motion of the gyro alone, not including the azimuthal motion.
This grid does not rotate with respect to observers at rest at spatial infinity. Along the geodesics $E_1$ rotates with respect to fixed Cartesian axes at radial infinity
(i.e., with respect to the Cartesian frame \eqref{cartesianframe})
 by a rate determined by the orbital angular velocity $d\phi/d\tau$ measuring the rate of rotation of these axes in the counterclockwise direction of increasing $\phi$ coordinate. Subtracting the angular velocity  ${\mathcal T}$ of the gyro axes in the clockwise direction gives the total coordinate time angular velocity of the gyro spin relative to axes whose directions are fixed along the orbit with respect to radial infinity as
\begin{eqnarray}\label{omegaprec}
  \Omega_{\rm (prec)} &=& \frac{d\phi}{d\tau} -\mathcal T  
  = \frac{d}{d\tau} \left(\phi-\Psi  \right)
\,.
\end{eqnarray}
This precession frequency corresponds to the counterclockwise rotation by $\phi$ of some initial spherical axes in the rest frame of the orbit (clockwise when $\phi<0$), choosing $\Psi=0$ at $\phi=0$ to align them with the spherical frame vectors initially at $\tau=0$,  which are then rotated clockwise by the angle $\Psi$ to keep them \lq\lq parallel" to the original axes (in the sense of parallel transport)
\beq
   \left(\begin{array}{c} e_{1}\\e_{3} \end{array} \right)
= \left(\begin{array}{cc} \cos(\phi-\Psi)&  \sin(\phi-\Psi)\\ -\sin(\phi-\Psi)&  \cos(\phi-\Psi) \end{array} \right)
   \left(\begin{array}{c}  e_{1}(0)\\  e_{3}(0) \end{array} \right)
\,.
\eeq
It is the difference between these two opposing angles which leads to a net precession with respect to the static observer grid.
Apart from spacetime relative tilting of the spin vector  which does not contribute to cumulative rotation of the spin (Wigner rotation effects \cite{Bini:2016iym}), this is just the rotation which leads to the spin angular velocity derived in Appendix A for a stationary asymptotically flat spacetime.

\begin{figure*}
\centering
\includegraphics[scale=0.35]{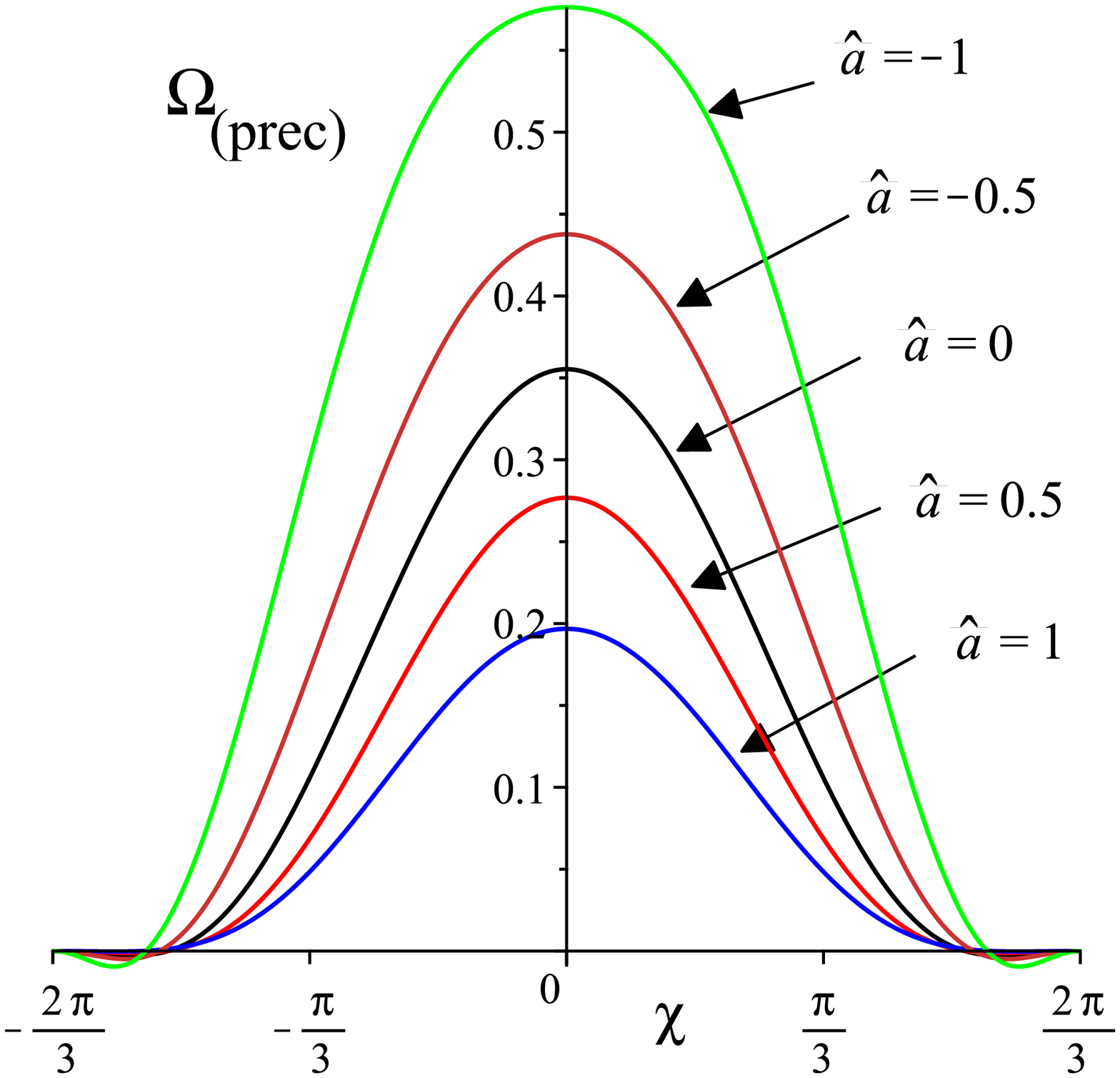}
\includegraphics[scale=0.35]{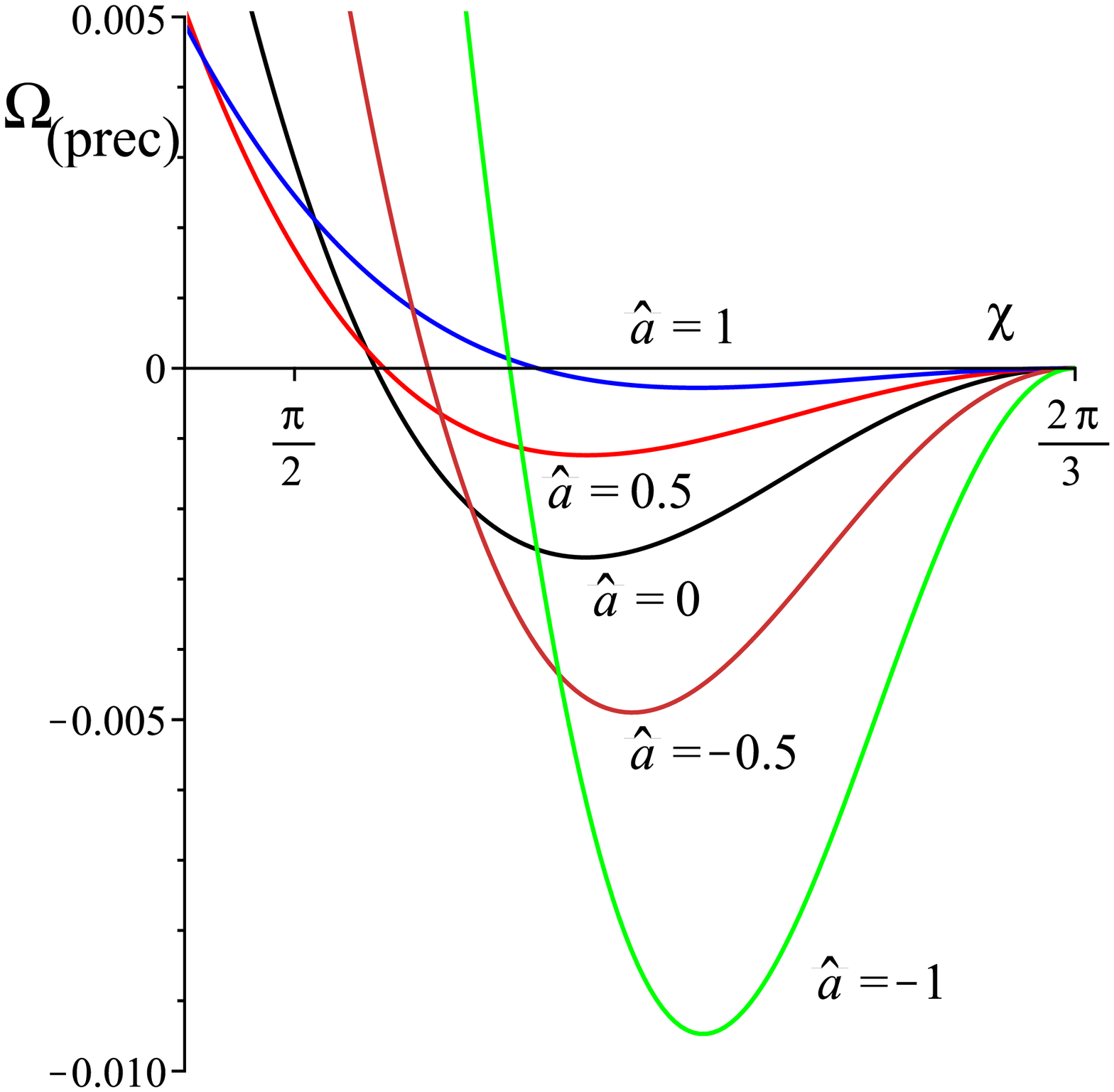}\\
\hbox to \hsize{ \hfil \kern 10pt $a)$ \hfil\hfil $b)$ \hfil}
\caption{\label{fig_omega}
The precession frequency $\Omega_{\rm (prec)} $ of a test gyroscope in Kerr
is shown as a function of $\chi$ for ${\hat a}=[0,0.5,1]$ for the prograde orbit of Fig.~\ref{orbits}$b$ ($e=2$ and $u_p=0.1$ and $L>0$).
The closeup plot $b)$ here  shows the negative values of $\Omega_{\rm (prec)} $, which vary on a much smaller scale compared to the positive values.
For the chosen parameter values, $\chi_{\rm max}=\arccos(-1/e)=2\pi/3=120^\circ$, while the zeros occur at $\chi_0\approx\pm 24.56^\circ$, corresponding to $r_0=11.2173M$ compared to the periastron $r_{\rm per}\approx 3.33 M$.}
\end{figure*}

\begin{figure*}
\centering
\includegraphics[scale=0.35]{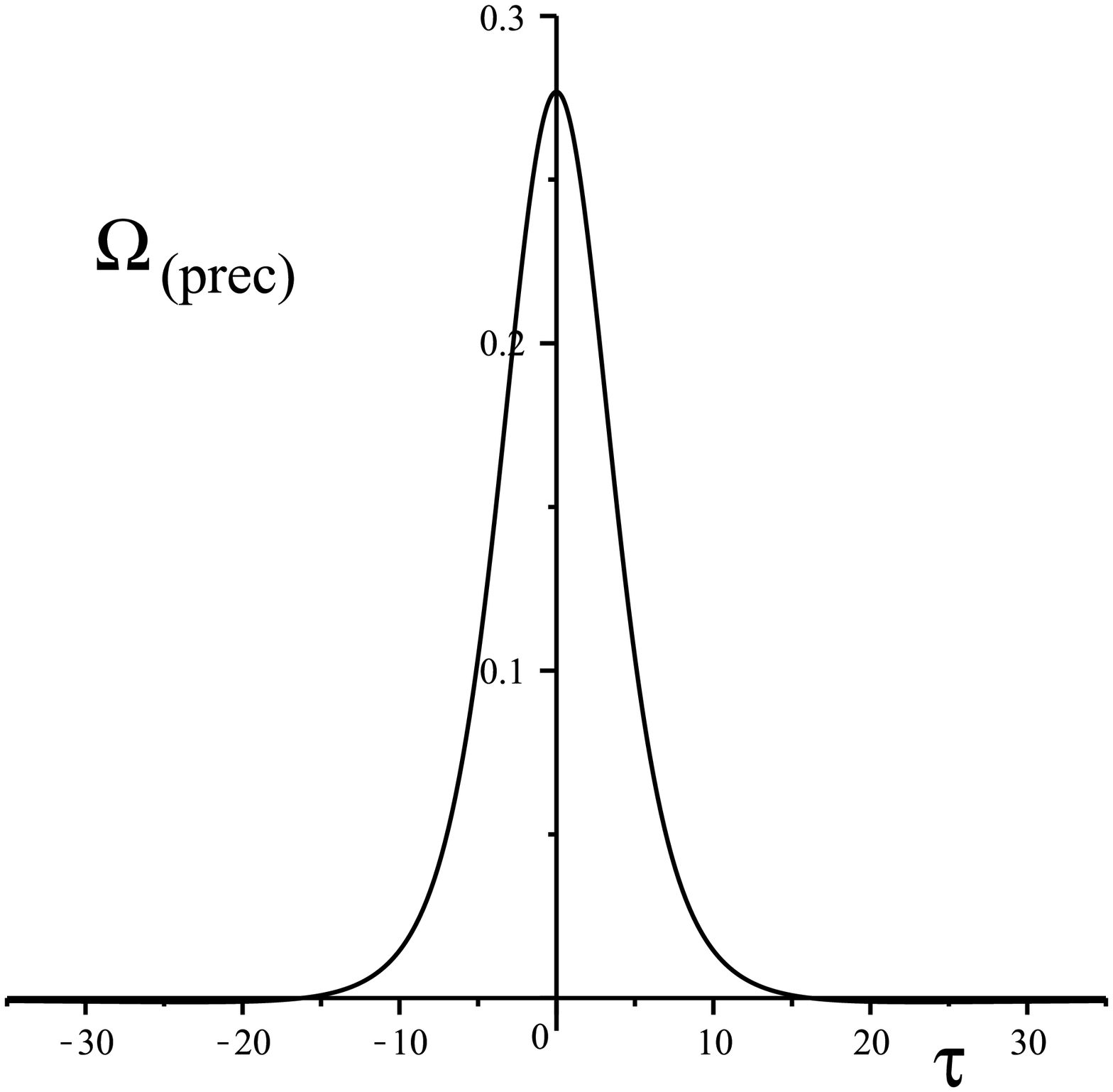}
\includegraphics[scale=0.35]{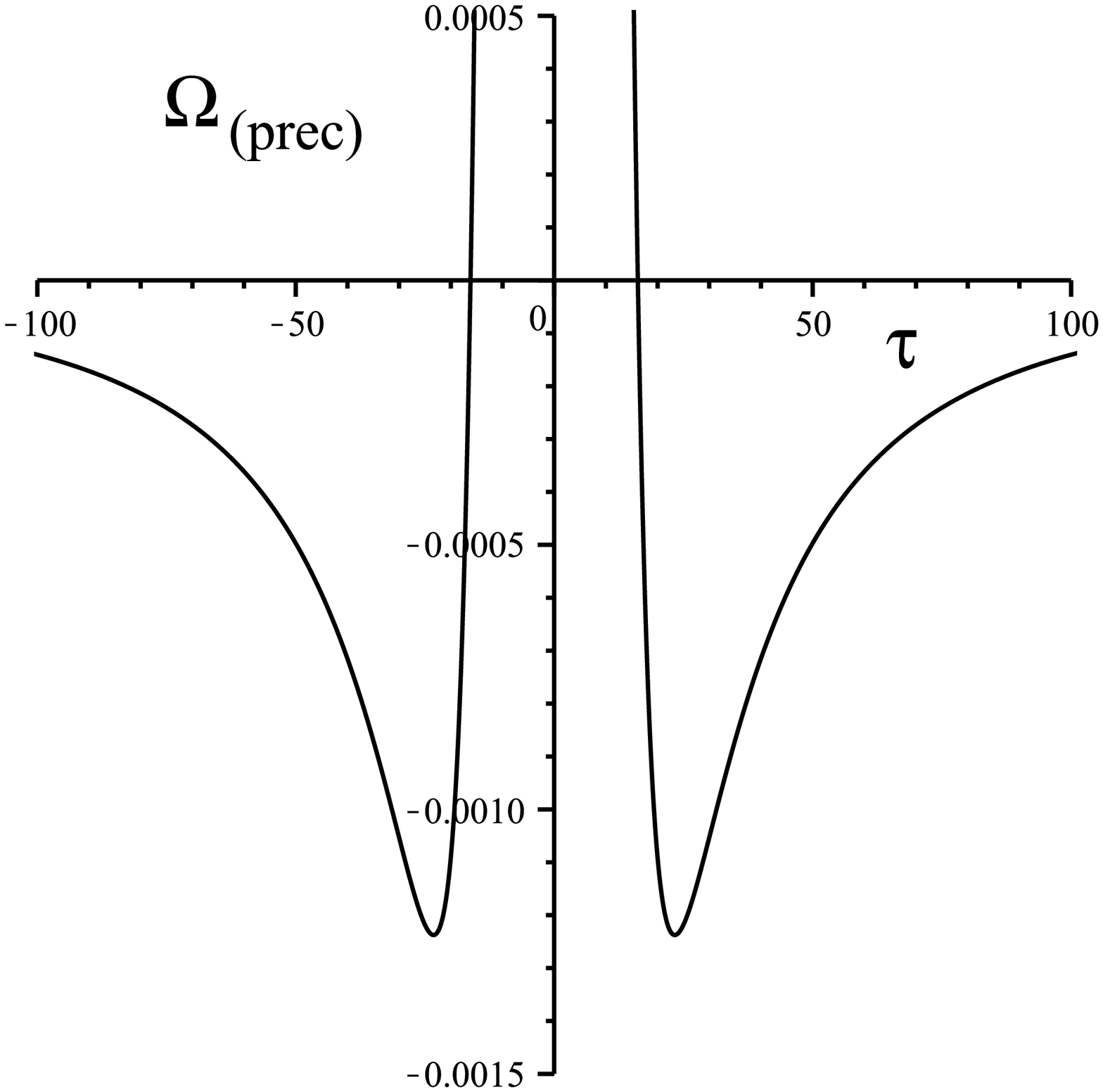}
\caption{\label{fig_omega_tau}
The precession frequency  $\Omega_{\rm (prec)} $ of a test gyroscope in Kerr  as a function of $\tau$ for the same parameter choice as in Fig.~\ref{fig_omega}, but with $\hat a=0.5$.
}
\end{figure*}

\begin{figure}
\centering
\includegraphics[scale=0.35]{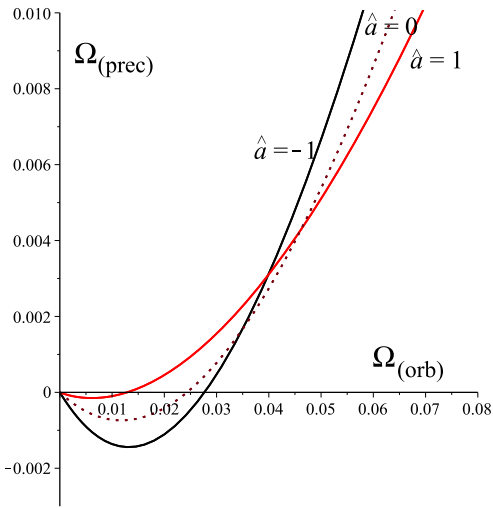}
\caption{
A plot of $\Omega_{\rm (prec)}$ versus $\Omega_{\rm (orb)}=d\phi/d\tau$
for $u_p=1/15$ and $e=2$ for selected values of $\hat a$. Positive values of $\Omega_{\rm (prec)}$ correspond to a forward precession (in the same sense as the orbital direction), and negative values to a backward precession. Note that the origin corresponds to $u=0$ ($r\to\infty$) where both frequencies go to zero.
}\label{figOmegaOmega}
\end{figure}

\begin{figure}
\includegraphics[scale=0.35]{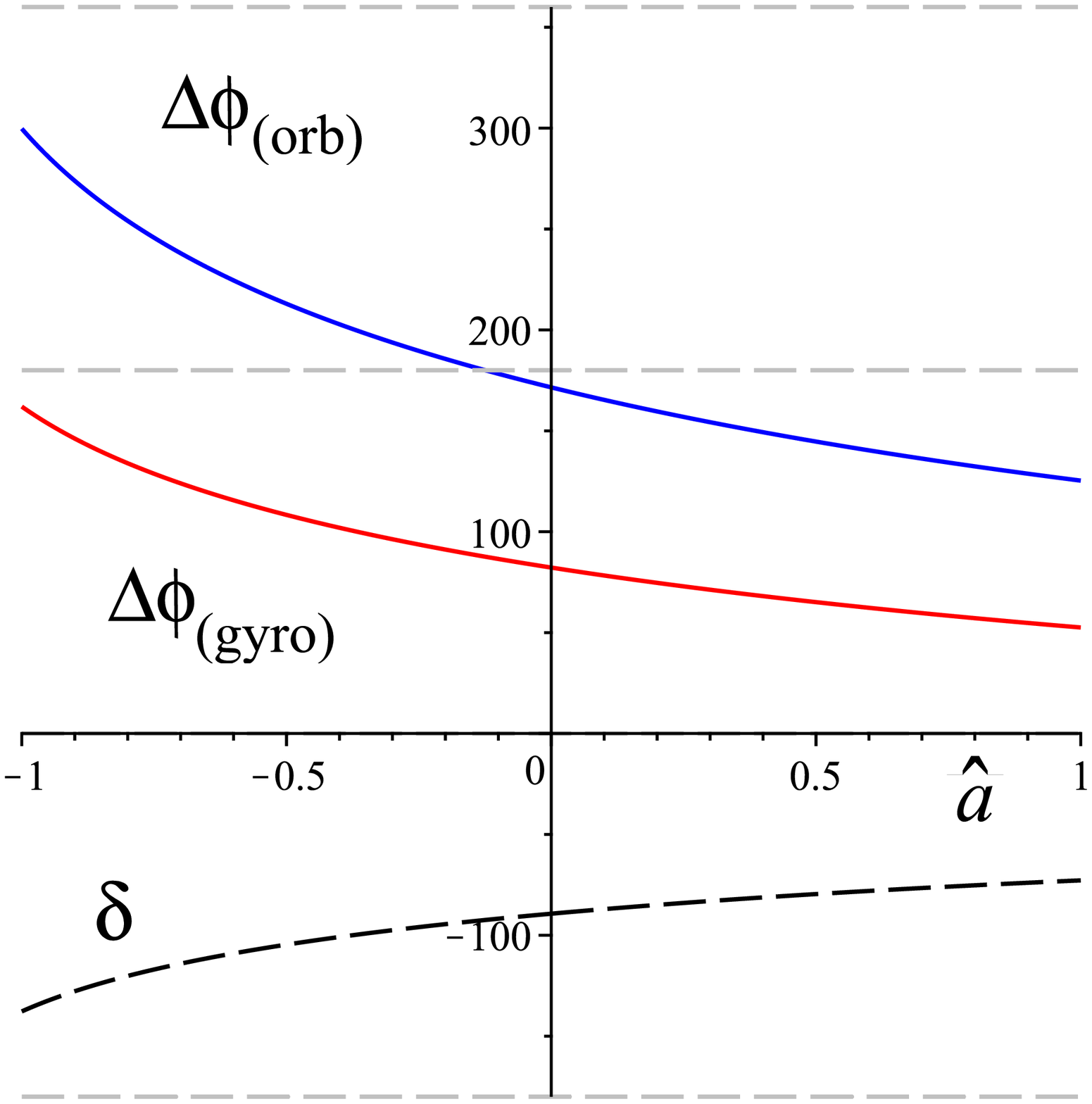}
\caption{\label{fig_delta_phis}
The total precession angle $\Delta\phi_{\rm(gyro)}$, the deflection angle $\Delta\phi_{\rm(orb)}$ and their difference $\delta$ (in degrees) for orbits with $\dot\phi>0$ are shown for $e=2$ and $u_p=1/15$ as functions of $\hat a$.
Dash-dotted horizontal lines are equally spaced by $180$ degrees.}
\end{figure}

Note that the proper time precession frequency has the following simple representation in terms of the constants of the motion
\beq
\label{prec-kerr}
\Omega_{\rm (prec)}  = \frac{L-2Mx/r}{\Delta}-\frac{a+E x}{r^2+x^2}\,.   
\eeq
Fig.~\ref{fig_omega} shows the typical behavior of $\Omega_{\rm (prec)} $ as an even function of $\chi$ for selected values of the black hole rotation parameter.
At $\chi=0, \pm \chi_{\rm (max)}$ the precession angular velocity has relative maxima, with relative minima in between.
Fig.~\ref{fig_omega_tau} shows instead the precession frequency as a function of the proper time along the orbit, exhibiting the same behavior.
The vanishing of the precession frequency occurs at a particular value of the radial coordinate along the orbit where the orbital frequency $\Omega_{\rm (orb)}\equiv d\phi/d\tau$ is balanced by the Frenet-Serret torsion $\mathcal T$.
Therefore, a positive precession frequency means orbital frequency dominance, whereas a negative one means Frenet-Serret torsion dominance (see Fig.~\ref{figOmegaOmega}).
The corresponding value of $r$ can be determined analytically by solving the cubic equation $\Omega_{\rm (prec)}=0$.
Finally, Fig.~\ref{fig_delta_phis} shows the total precession angle 
\beq
\Delta\phi_{\rm(gyro)}=2\int_{0}^{\chi_{\rm(max)}}\Omega_{\rm(prec)}\frac{d\tau}{d\chi}d\chi\,,
\eeq
the deflection angle 
\beq
\Delta\phi_{\rm(orb)}=2\int_{0}^{\chi_{\rm(max)}}\frac{d\phi}{d\chi}d\chi-\pi\,,
\eeq
and their difference $\delta=\Delta\phi_{\rm(gyro)}-\Delta\phi_{\rm(orb)}$ as functions of $\hat a$.
Indeed, explicit formulas for these quantities are not very illuminating. 
Note that restrictions to the allowed values of the eccentricity arise from the existence of the separatrix curve in the $(u_p,e)$ space. For example, for $\hat a=0.1$ and $u_p=0.05$ it turns out that $e\leq e^{\rm sep}\approx7.2660$.

 
\subsection{The Schwarzschild limit}

Since the general formulas describing this problem are rather complex for nonzero rotation, it is instructive to consider the Schwarzschild limit $a=0$ where reasonable formulas can be found.

Along the orbit the energy and angular momentum (per unit mass) are constant and are related to the eccentricity and the semilatus rectum $p=1/u_p$ by
\begin{eqnarray}\label{ELep}
E^2&=&\frac{(1-2u_p)^2-4u_p^2e^2}{1-3u_p-u_pe^2}\,,\nonumber\\
\frac{L^2}{M^2}&=&\frac{1}{u_p(1-3u_p-u_pe^2)}\,,
\end{eqnarray}
where now $E\ge1$ (and also $e\ge1)$, with equality holding for the parabolic-like orbits.
[The expression for $E$ in terms of $L$ requires the solution of a cubic equation.] 

Using $\chi$ in place of $r$, the $t$- and $\phi$-geodesic equations \eqref{dtdchi_dphidchi} for prograde orbits become
\begin{eqnarray}
\frac{d\chi}{d\tau}&=& \frac{u_p^{3/2}}{M}\frac{(1+e\cos\chi)^2 [1-6u_p-2eu_p\cos \chi]^{1/2}}{(1-3u_p-u_pe^2)^{1/2}}\,,\nonumber\\
\frac{d\phi}{d\chi}&=& \frac{1}{[1-6u_p-2eu_p\cos\chi]^{1/2}}\,.
\end{eqnarray}
The latter equation can be exactly integrated, choosing $\phi(0)=0$, to yield \cite{Chandrasekhar:1985kt,Bini:2016ubc}
\beq
\label{phi_sol_elliptic}
\phi(\chi) =\frac{m}{\sqrt{eu_p}}\left[ K(m)-F\left(\cos\frac{\chi}{2},m\right) \right]\,,
\eeq
where we recall
\beq
\label{m_def}
m=2 \sqrt{\frac{eu_p}{1-6u_p +2eu_p}}\,.
\eeq
This solution coincides with Eq. \eqref{phiuschw} when expressing $u$ in terms of $\chi$ as $u=u_p(1+e\cos\chi)$.
The above representation in terms of elliptic integrals is valid only for the range of values $0<m<1$.
The limiting value  $m= 1$ where it fails corresponds to the relationship 
\beq
e =\frac{1-6 u_p}{2 u_p}=\frac{p-6}{2}\quad
\rightarrow\quad p\ge8
\,,
\eeq
which describes the separatrix \eqref{separatrix} for $a=0$ beyond which these orbits are captured by the black hole.
For example, $m=1$ when $u_p=0.1$ and $e=2$, while $r_{\rm(per)}/M=10/3$.

Finally, the total deflection angle over the entire orbit is given by
\beq
\Delta\phi_{\rm(orb)}=2\phi(\chi_{\rm (max)})-\pi\,,
\eeq
where $\chi_{\rm (max)}=\arccos(-1/e)$ and
\beq
\phi(\chi_{\rm (max)})=\frac{m}{\sqrt{eu_p}}\left[ K(m)-F\left(\sqrt{\frac{e-1}{2e}},m  \right) \right]\,,
\eeq
in agreement with Eq. \eqref{phi0schw}, since $\chi=\chi_{\rm (max)}$ corresponds to $u=0$.
On the other hand the precession of the hyperbolic orbit can be inferred as the difference between the increment of $\phi$ compared to the increment of $\chi$ for the Newtonian orbit with these parameters, namely $2(\phi(\chi_{\rm (max)})-\chi_{\rm (max)})$, as shown in Fig.~\ref{fig_bob}.

The gyroscope angular velocity \eqref{prec-kerr} simplifies to
\begin{eqnarray}\label{prec-schwarz}
\Omega_{\rm (prec)} &=&\frac{L}{r^2}\left(1-\frac{E}{1+L^2/r^2}\right)
\end{eqnarray}
and behaves like ${(1-E)L}/{r^2} $ as $r\to\infty$ (and $\chi\to\pm\chi_{\rm (max)}$), which has the opposite sign of $L$ for hyperbolic-like unbound orbits with $E>1$ in that limit, corresponding to a rotation in the opposite sense compared to the azimuthal motion, but vanishing in the limit of large radii.
This sign changes within the radius
$r_0=\sqrt{L/(E-1)}\equiv M/u_0$ where the angular velocity is much larger and has the same sign as for bound orbits with $E<1$. This change occurs at the two values of $\chi$ defined by
\beq
\cos \chi_0 =\frac{1}{e}\left( \frac{u_0}{u_p}-1 \right)\,.
\eeq
At the  periastron, $\chi=0$, the precession frequency is given by
\begin{eqnarray}
&&\Omega_{\rm (prec)} \vert_{\chi=0}
\nonumber\\ && 
 =Lu_p^2 (1+e)^2  \left(1-\frac{E}{1+L^2u_p^2 (1+e)^2}\right)\,,
\end{eqnarray}
which is also an absolute maximum.
The derivative $d\Omega_{\rm (prec)} /dr$ then vanishes at  
\beq
r=\frac{|L|\sqrt{1+\sqrt{E}}}{\sqrt{E-1}}=r_0 \sqrt{1+ \sqrt{E}}\,,
\eeq
which for positive $L$ corresponds to the relative minima of Fig.~\ref{fig_omega}.

To come full circle in this discussion of the unified nature of Thomas precession, Wigner rotations and variable direction boosts within the Lorentz group, return for a moment to the case of bound circular orbits in the Schwarzschild spacetime, where the formula \eqref{prec-schwarz} reproduces exactly the classical Thomas precession formula for an accelerated circular orbit in flat spacetime given by Eq.~(54) of Ref.~\cite{Bini:2002mh}.
To see this start from the values of $E$ and $L$ for circular orbits (see Eqs.~\eqref{ELep} with $e=0$ and $u_p=u=M/r=$ const.) 
\beq
E = \frac{1-2u}{\sqrt{1-3u}}\,,\qquad
L= \frac{M}{\sqrt{u(1-3u)}}\,,
\eeq
and use these to rewrite the equations of motion \eqref{dtdchi_dphidchi} as
\begin{eqnarray}
\frac{dt}{d\tau} &=& \frac{E}{r(r-2M)}=  \frac{1}{\sqrt{1-3u}}
\equiv \Gamma\,,
\nonumber\\
\frac{d\phi}{d\tau} &=& \frac{L}{r^2}=  \frac{\Gamma u^{3/2}}{M}
\equiv \omega_{\rm{(orb,\tau)}}
\,,
\end{eqnarray}
or equivalently
\beq
\frac{d\phi}{dt}=\frac{u^{3/2}}{M}\equiv \omega_{\rm{(orb,t)}}
= \Gamma^{-1} \omega_{\rm{(orb,\tau)}}
 \,.
\eeq
Similarly one finds
\begin{eqnarray}
\frac{d\Psi}{d\tau}
&=&
E L/(r^2+L^2) = \frac{d\phi}{dt}
\,.
\end{eqnarray}
Then the precession formula becomes 
\begin{eqnarray}
\Omega_{\rm(prec)}
&=& \frac{d\phi}{d\tau} -\frac{d\Psi}{d\tau}
\nonumber\\
&=& (\Gamma -1) \,\omega_{\rm(orb,t)}
\nonumber\\
&=& (1-\Gamma^{-1}) \,\omega_{\rm(orb,\tau)}
\nonumber\\
&=& \omega_{\rm(orb,\tau)} - \omega_{\rm(orb,t)}
\,.
\end{eqnarray}
In other words the precession is a simple mismatch of the forward orbital angular velocity $d\phi/d\tau$ of the spherical axes and the backward angular velocity $d\Psi/d\tau= d\phi/dt$
of the parallel transported axes to compensate. These are nearly the same but require an additional gamma factor on one or the other to compare to each other on the same clock scale, resulting in their failure to exactly cancel. The net result is a forward rotation of the spin vector undergoing this transport.
Orbit eccentricity and rotation of the spacetime certainly complicate matters, but the basic point of departure is this simple fact that goes all the way back to the example of a classical electron in flat spacetime.

\section{Concluding remarks}

The precession of the spin vector of a test gyroscope in geodesic motion in a Kerr spacetime following a precessing (no capture) conic section orbit  has been extended from the bound to the unbound case (parabolic and hyperbolic-like orbits), generalizing previous results~\cite{Bini:2016iym}.
The spin of a gyroscope undergoes parallel transport along the orbit, and its precession is measured with respect to a static frame whose axes are aligned with the fixed stars at spatial infinity. In the construction of such a frame a key role is played by an intermediate degenerate Frenet-Serret frame first introduced by Marck~\cite{marck1,marck2}, which is then suitably rotated to achieve
parallel transport.
The advantage of using Marck's frame locked to the geometry of the orbit is that the analysis is significantly simplified.

We have computed the precession frequency and the corresponding total precession angle after a complete scattering process coming in from radial infinity to the periastron and then returning to radial infinity,
both in the rotating Kerr case and in its non-rotating Schwarzschild limit.  
Due to the symmetry of this process with respect to the periastron, the precession frequency is a symmetric function of the radial variable, reaching its maximum value (depending on the rotational parameter of the hole as well as on the orbital parameters, i.e., eccentricity and semi-latus rectum) at the minimum approach distance. While positive near the black hole, the precessional rotation of the spin is in the same sense as the orbital rotation, but 
as the radial variable increases, the precession frequency decreases to zero and then becomes negative, reversing direction.
This reversal occurs when the orbital frequency equals the Frenet-Serret torsion of the orbit, and can be determined analytically. Indeed the spin precession is a direct competition between the orbital angular velocity and Frenet-Serret torsion of the orbit in which the latter wins out far enough from the horizon.
Known results for both bound and circular orbits in the equatorial plane have been re-obtained as a consistency check. 

A number of geometrical features associated with these two components have been elucidated with an appendix devoted to the study of their relationship to the classical Thomas precession and variable boost geometry of the Lorentz group, in turn flowing from the Wigner rotation property of successive boosts within that group.
We have also discussed there the further decomposition of the gyroscope precession formula in terms of the spatial geometry precession frequency and Fermi-Walker transport frequency revisiting previous results and throwing new light on them.

Finally, the situation considered here is interesting as a preliminary analysis for computing gravitational self-force/self-torque effects associated with hyperbolic encounters between black holes.
In fact, one expects an enhanced gravitational wave emission at periastron, which can be detected by the current capability of detectors.
The main difficulty at the moment in any modeling of such processes is to account for the continuous spectrum of the emission. Valuable attempts use the ``effective-one-body'' model as in Ref.~\cite{Damour:2014afa} and post-Newtonian theory, but so far have no counterpart in black hole perturbation theory.
This will be a challenge for future work.

\appendix

\section{Spin precession in stationary asymptotically flat spacetimes}

For a timelike world line $x(\tau)$ parametrized by the proper time $\tau$, the intrinsic covariant derivative along the world line 
$D(U) X/d\tau$
coincides with the covariant derivative $\nabla_{\hbox{U}}$ when acting on tensor fields on the spacetime 
\beq
\frac{D(U)}{d\tau} (T\circ x(\tau)) = \nabla_{\hbox{$U$}} T\circ x(\tau)\,.
\eeq
From now on the restriction ``$\circ\, x(\tau)$" will be understood when needed and not written explicitly.
The world line is in general not geodesic, i.e., with nonzero acceleration $a(U)=DU/d\tau$.

Let $D{}_{\rm(fw)}(U,u) T/d\tau = P(u)D(U) T/d\tau$ for any tensor $T$ defined along the world line, and hence for a tensor field $T$ defined on the spacetime 
\begin{eqnarray}\label{Dfw}
\frac{D{}_{\rm(fw)}(U,u)}{d\tau} T &=& \gamma(U,u) P(u) \nabla_{\hbox{$u+\nu(U,u)$}} T
\nonumber\\
&=& \gamma(U,u) [P(u) \nabla_{\hbox{$u$}} 
          +P(u) \nabla_{\hbox{$\nu(U,u)$}}] T
\nonumber\\
&\equiv& \gamma(U,u) [\nabla{}_{\rm(fw)}(u) 
          + \nabla(u)_{\hbox{$\nu(U,u)$}}] T
\,,\nonumber\\
\end{eqnarray}
which defines the spatial Fermi-Walker derivative $\nabla{}_{\rm(fw)}(u)$ along $u$ and the spatial covariant derivative $\nabla(u) $ orthogonal to $u$.

Fermi-Walker transport of a vector $X$ orthogonal to $U$ along the world line satisfies $P(U) D(U) X/d\tau = 0$.
The Fermi-Walker transport of the spin vector $S\in LRS_U$ along $U$ therefore takes the form
\beq
  B{}_{\rm(lrs)}(u,U) \frac{D(U)S}{d\tau} =P(u) B(u,U) P(U)\frac{D(U)S}{d\tau} =0\,.
\eeq
The boosted spin vector into $LRS_u$ is 
\beq
\mathcal{S}(U,u) = B(u,U)S=B{}_{\rm(lrs)}(u,U) S\,,
\eeq 
so its derivative is 
\begin{widetext}
\begin{eqnarray}\label{DS}
\frac{D{}_{\rm(fw)}(U,u)}{d\tau} \mathcal{S}(u,U)
&=& P(u) \frac{D(U)}{d\tau} (B{}_{\rm(lrs)}(u,U) S)
\nonumber\\
&&=P(u) \left[\frac{D(U)}{d\tau} B{}_{\rm(lrs)}(u,U) \right] S
+ P(u) B{}_{\rm(lrs)}(u,U) \left[\frac{D(U)}{d\tau} S \right] 
\nonumber\\
&&= \left[\frac{D{}_{\rm(fw)}(U,u)B{}_{\rm(lrs)}(u,U)}{d\tau} B{}_{\rm(lrs)}(u,U)^{-1}\right]\mathcal{S}(U,u)
\nonumber\\
 &&= \gamma(U,u)\zeta{}_{\rm(fw)}(U,u) \times_u \mathcal{S}(U,u) \,,
\end{eqnarray}
\end{widetext}
where $\zeta{}_{\rm(fw)}(U,u)$ defines the Fermi-Walker angular velocity of the boosted spin vector with respect to $u$.
Thus the relative angular velocity of the spin with respect to the static frame is entirely due to the natural group theoretical ``logarithmic" derivative of the relative boost, which defines an antisymmetric linear transformation generating a rotation along the world line.

One finds  that the Fermi-Walker angular velocity is the sum of two terms: a \lq\lq geodesic precession" term and a \lq\lq Thomas precession" term \cite{Jantzen:1992rg}, namely
\beq
\zeta{}_{\rm(fw)}(U,u)=\zeta_{\rm(geo)}(U,u)+\zeta_{\rm(thom)}(U,u) 
\eeq
with
\begin{eqnarray}
\zeta_{\rm(geo)}(U,u)&=& \frac{1}{\gamma(U,u)   + 1 } \, \nu(U,u) \times_u  F{}_{\rm(fw)}^{\rm(G)}(U,u) \nonumber\\
\zeta_{\rm(thom)}(U,u) &=&  - \frac{\gamma(U,u)}{   \gamma(U,u)   +1}      \nu(U,u) \times_u P(u)a(U)\,,\nonumber\\
\end{eqnarray}
where the \lq\lq gravitational force" entering the geodesic precession frequency is defined by
\beq
F{}_{\rm(fw)}^{\rm(G)}(U,u)=-\frac{D{}_{\rm(fw)}(U,u)}{d\tau}\,u\,.
\eeq
We find
\begin{eqnarray}
F{}_{\rm(fw)}^{\rm(G)}(U,u)=-\frac{M}{r^3}\frac{(Er^2-ax)}{N \sqrt{\Delta}}E_1-\frac{aMU^r}{r^2 N^2 \sqrt{\Delta}}E_3\,,
\end{eqnarray}
which in the Schwarzschild case reduces to
\begin{eqnarray}
F{}_{\rm(fw)}^{\rm(G)}(U,u)=-\frac{M}{r^2}\frac{E }{ N^2  }E_1=-\gamma\frac{M}{r^2 N }\, E_1\,.
\end{eqnarray}

This ``relative gravitational force" involving the acceleration and vorticity of $u$ has been represented terms of \lq\lq gravitoelectric" and ``gravitomagnetic" fields  in analogy with the electromagnetic Lorentz force in \cite{Jantzen:1992rg}.
The Thomas precession term vanishes for geodesic motion $a(U)=0$ and generalizes the special relativistic Thomas precession associated with accelerated motion with respect to an inertial frame, leaving only the first term to characterize spin precession along a geodesic, which is a gravitational analog of the Thomas precession due to the relative gravitational force. 
The geodesic precession is then given by $\zeta{}_{\rm(fw)}=\zeta{}_{\rm(fw)}^2E_2$, with
\beq
\zeta{}_{\rm(fw)}^2=-\frac{M}{r^3}\frac{a+Ex}{E(E+N)}\,,
\eeq
which becomes 
\beq
\zeta{}_{\rm(fw)}^2=-\frac{M}{r^3}\frac{L}{E+N}\,,
\eeq
in the Schwarzschild case.

Summarizing, the spin vector $S\in LRS_U$ of a gyroscope carried along a world line with 4-velocity $U$ undergoes Fermi-Walker transport along $U$,
\beq
 P(U)\frac{D}{d\tau}S=0\,.
\eeq
Given any family of test observers with 4-velocity $u$ defined in some open tube around the gyro world line, one can consider the orthogonally projected relative version of this transport law with respect to $u$.
We have shown that the boosted spin vector ${\mathcal S}(U,u)=B_{\rm (lrs)}(u,U)S$  onto $LRS_u$ (abbreviated by $\mathcal{S}$ below) satisfies the following transport equation
\beq
\label{eq_of_the_boosted_spin_vector}
\frac{D{}_{\rm(fw)}(U,u)}{d\tau} \mathcal{S}=\gamma(U,u) \,\zeta{}_{\rm(fw)}(U,u)\times_u \mathcal{S}\,.
\eeq
Now consider the component form of these equations with respect to various orthonormal frames defined all along $U$ adapted either to an orthogonal decomposition with respect to $U$ or to $u$. We examine the following cases:

\begin{enumerate}
  \item Frame adapted to $U$ and Fermi-Walker transported along $U$.\\
Let $e_i\in LRS_U$ be a spatial triad orthogonal to $U$ and Fermi-Walker transported along $U$, $P(U)De_i/d\tau=0$.
Then $S=S^i e_i$ and 
\beq
\frac{dS^i}{d\tau}=0\,,
\eeq
i.e., the components of the spin vector with respect to this frame are constant.

  \item Frenet-Serret frame for $U$.\\
Let $\tilde e_i$ be the spatial triad orthogonal to $U$ which together with $U$ satisfies the spatial Frenet-Serret relations, 
\beq
P(U)\frac{D}{d\tau}\tilde e_i=\omega_{\rm FS}\times_U \tilde e_i \,,
\eeq
with $\omega_{\rm FS}=-{\mathcal T}E_2$ (see Eq.~\eqref{K31bis}).
Then $S=\tilde S^i \tilde e_i$ and letting  $\eta(U)_{ijk}$ be the unit 3-form components in this frame,
\beq
\frac{d \tilde S^i}{d\tau }  = -[\omega_{\rm FS}\times_U S]^i
\equiv -\eta(U)^i{}_{jk}\omega_{\rm FS}^j \tilde S^k
\,.
\eeq

  \item Frame adapted to $u$ obtained by boosting onto $LRS_u$ a Frenet-Serret frame along $U$.\\
A spatial Frenet-Serret frame $\tilde e_i\in LRS_U$ can be boosted to a frame ${\mathcal E}_i=B_{\rm (lrs)}(u,U)\tilde e_i\in LRS_u$ given by
\beq
{\mathcal E}_i=\tilde e_i +\frac{U+u}{\gamma +1} (u\cdot \tilde e_i)\,,
\eeq
with the abbreviation  $\gamma=\gamma(U,u)=-U\cdot u$.
The boosted spin vector  ${\mathcal S}\in LRS_u$ (undergoing the transport law \eqref{eq_of_the_boosted_spin_vector}) then has components
\begin{eqnarray}
{\mathcal S}&=& B_{\rm (lrs)}(u,U)S
= \tilde S^i {\mathcal E}_i
\,,
\end{eqnarray}
so that by the definition \eqref{Dfw} one has
\beq
\frac{d \tilde S^i}{d\tau} {\mathcal E}_i+\tilde S^i P(u)\nabla_U {\mathcal E}_i
=\gamma\, \zeta{}_{\rm(fw)} \times_u {\mathcal S}\,.
\eeq
Introduce the frame transport angular velocity $\Omega_{\mathcal E}$ by
\beq
P(u)\nabla_U {\mathcal E}_i=-\gamma\, \Omega_{\mathcal E} \times_u {\mathcal E}_i\,.
\eeq
We then have 
\begin{eqnarray}
&&\label{S_in_u}
\frac{d \tilde S^i}{d\tau} 
=\gamma\, [(\zeta{}_{\rm(fw)}+\Omega_{\mathcal E}) \times_u {\mathcal S}]^i
\nonumber\\
&&\ =\gamma\, \eta(u)^i{}_{jk}[\zeta{}_{\rm(fw)}^j+\Omega_{\mathcal E}^j]\, {\mathcal S}^k
\,.
\end{eqnarray}
In the present case we have $\Omega_{\mathcal E}=\Omega_{\mathcal E}^2{\mathcal E}_2$ with
\beq
\gamma \Omega_{\mathcal E}^2= (a+xE)\left(\frac{1}{r^2+x^2}+\frac{M}{r^3 } \frac{1}{N(E+N)} \right)\,.
\eeq
Equivalently,
\beq
\gamma \Omega_{\mathcal E}^2-{\mathcal T}=\frac{M}{r^3 } \frac{(a+xE)}{N(E+N)}
=-\gamma\zeta{}_{\rm(fw)}^2\,.
\eeq
In the Schwarzschild case, with $E=\gamma N$ we have
\begin{eqnarray}
\gamma \Omega_{\mathcal E}^2-{\mathcal T} &=& \frac{M}{r^3 } \frac{ LE }{N(E+N)}\,.
\end{eqnarray}

  \item Static frame adapted to $u$, restricted  along $U$.\\
Let $E_i$ complete $u=(-g_{tt})^{-1/2}\partial_t$  to a static orthonormal frame, therefore satisfying $P(u)\pounds_{\hbox{$\partial_t$}} E_i=0$, the condition which rigidly ties the frame to the static coordinate grid which in turn is nonrotating at spatial infinity, and let
\beq
{\mathcal S}={\mathcal S}^i E_i\,.
\eeq
Introduce the frame transport angular velocity $\Omega_{E}$ along $U$,
\beq
P(u)\nabla_U E_i=-\gamma\, \Omega_{E} \times_u E_i\,.
\eeq
We then have 
\beq
\frac{d  {\mathcal S}^i}{d\tau} 
=\gamma\, [(\zeta{}_{\rm(fw)}+\Omega_E) \times_u {\mathcal S}]^i\,.
\eeq

In the present case we have $ \Omega_E = \Omega_E^2 E_2$ with
\begin{eqnarray}
\gamma \Omega_E^2 
&=& \frac{aE(1+N^2)+2x N^2}{2N\Delta}-x\frac{Ma^2}{r^3N\Delta} \nonumber
\end{eqnarray}
In the Schwarzschild limit
\begin{eqnarray}
\gamma \Omega_E^2 
&=& \frac{L }{r^2 N} \,.
\end{eqnarray}
\end{enumerate}

Note that the two frames in $LRS_u$, ${\mathcal E}_i$ and the static frame $E_i$ are related by a (spatial) rotation $E_i =\mathcal{E}_j R^j{}_i$
 by a counterclockwise angle $\Lambda$ in $LRS_u$, namely (since $E_2={\mathcal E}_2$ is unchanged)
\begin{eqnarray}
E_1 &=& \cos \Lambda {\mathcal E}_1+\sin \Lambda {\mathcal E}_3\nonumber\\
E_3 &=& -\sin \Lambda {\mathcal E}_1+\cos \Lambda {\mathcal E}_3\,,
\end{eqnarray} 
that is
\begin{eqnarray}
   \begin{pmatrix} E_1 & E_2 & E_3 \end{pmatrix}
&=&  \begin{pmatrix} \mathcal{E}_1 & \mathcal{E}_2 & \mathcal{E}_3 \end{pmatrix} \,
    \begin{pmatrix} \cos\Lambda & 0 & -\sin\Lambda\\ 
		                   0 & 1 & 0\\
                    \sin\Lambda & 0 & \cos\Lambda
      \end{pmatrix}  \,.\nonumber
\end{eqnarray}

This is the same rotation $R(\Lambda)$ as the one in $LRS_U$ relating instead the static frame boosted to $LRS_U$ to the preliminary Marck frame $\tilde e_i$ since the boost is an isometry. The angle of the latter rotation was explicitly evaluated in Eqs.~C.15--17 of Ref.~\cite{Bini:2016iym} in terms of the relative velocity of the gyro, and represents a non-accumulating wobble of the boosted spin vector, so we can ignore its contribution to the precession of the spin in calculating the total rotation of the gyro spin over its unbounded orbit.
The associated relative angular velocity of the two spatial frames is defined by
\beq
[\dot R\, R^{-1}]^i{}_k=\gamma\, \eta(u)^i{}_{km}\dot\Lambda^m\,,
\eeq
which upon comparison of the two equations for the spin components must satisfy
\beq
\dot \Lambda = \Omega_{E}-\Omega_{\mathcal E}\,. 
\eeq

Finally one can evaluate the angular velocity of the static frame in terms of the vorticity of $u$ and the connection components of the frame
\begin{eqnarray}
  \nabla{}_{\rm(fw)}(u) E_i &=& -\eta(u)^{kj}{}_i \omega(u)_k E_j
\,,\nonumber\\
&=& \vec \omega(u)\times_m E_i\,,
\end{eqnarray}
where $\vec\omega(u)$ is given in Eq. \eqref{eq_a_omega}, $\vec\omega(u) =  -\frac{aM}{r^3N^2} \,E_2$.
In fact
\beq
\vec \omega(u)\times_m E_i=\omega(u)^j E_j \times_m E_i=\omega(u)^j \eta(u)_{ji}{}^k E_k\,.
\eeq
Splitting the covariant derivative along $U=\gamma (u+\nu)$
\beq
   P(u)\nabla_U E_i = \gamma\, [\nabla_{\rm(fw)}(u) E_i
 +P(u)\nabla_{\hbox{$\nu$}}\, E_i]
\eeq
one identifies a \lq\lq spatial curvature" angular velocity
\beq
\gamma P(u)\nabla_{\hbox{$\nu$}}\, E_i = \gamma \zeta_{\rm sc}\times_u E_i\,, 
\eeq
such that $\zeta_{\rm sc}=\zeta_{\rm sc}^2 \, E_2$, with
\beq
\zeta_{\rm sc}^2 = -\frac{r^3N^4-Ma^2}{r^3 N^2 \sqrt{\Delta}}\nu^3\,.
\eeq
We have then
\beq
   \mathcal{S}^i P(u)\nabla_U E_i 
= \gamma\, [\vec\omega(u)\times_u \mathcal{S} + \Gamma^k{}_{ji}\nu^j \mathcal{S}^i \, E_k]
\eeq
and
\beq
\Omega_{E}^i =\omega(u)^i +\zeta_{\rm sc}^i
\,.
\eeq
The first 
term is the ``gravitomagnetic" contribution to the angular velocity and the second is the ``space curvature" contribution \cite{Jantzen:1992rg}.

\subsection*{Acknowledgments}
D.B. thanks the Italian INFN (Naples) for partial support.
All  the authors are grateful to the International Center for Relativistic Astrophysics Network based in Pescara, Italy for partial support.


\begin{thebibliography}{00}

\bibitem{thirring1}
H. Thirring,
``\"Uber die Wirkung rotierender ferner Massen in der Einsteinschen Gravitationstheorie (On the Effect of Rotating Distant Masses in Einstein's Theory of Gravitation)," 
Phys.\ Zeit.\ {\bf 19}, 33 (1918). 

\bibitem{thirring2}
H. Thirring, 
``Berichtigung zu meiner Arbeit: `Über die Wirkung rotierender Massen in der Einsteinschen Gravitationstheorie' (Correction to my paper "On the Effect of Rotating Distant Masses in Einstein's Theory of Gravitation"),"
Phys.\ Zeit.\ {\bf 22}, 29 (1921).
 
\bibitem{lense-thirring}
J. Lense and H. Thirring,
``Über den Einfluss der Eigenrotation der Zentralkörper auf die Bewegung der Planeten und Monde nach der Einsteinschen Gravitationstheorie (On the Influence of the Proper Rotation of Central Bodies on the Motions of Planets and Moons According to Einstein's Theory of Gravitation),"
Phys.\ Zeit.\ {\bf 19}, 156 (1918).


\bibitem{schiff}
L.~I. Schiff 
``Motion of a gyroscope according to Einstein's theory of gravitation,"
Proc.\ Nat.\  Acad.\ Sci.\ {\bf 46}, 871 (1960).

\bibitem{gpb1}
See the GPB website at \url{https://einstein.stanford.edu}.

\bibitem{gpb2}
C.~W.~F. Everitt, et al, 
``Focus issue: Gravity Probe B,"
Class.\  Quant.\ Grav.\ {\bf 32} (2015).


\bibitem{RindlerPerlick:1990}
W. Rindler and V. Perlick,
``Rotating coordinates as tools for calculating circular geodesics and gyroscopic precession,"
Gen.\ Relativ.\ Grav.\ {\bf22}, 1067 (1990);
\lq\lq Erratum," Gen.\ Relativ.\ Grav.\ {\bf23}, 119 (1991).

\bibitem{Iyer:1993qa} 
  B.~R.~Iyer and C.~V.~Vishveshwara,
  ``The Frenet-Serret description of gyroscopic precession,''
  Phys.\ Rev.\ D {\bf 48}, 5706 (1993)
  [gr-qc/9310019].
	
\bibitem{Bini:1997eb} 
  D.~Bini, P.~Carini and R.~T.~Jantzen,
  ``The intrinsic derivative and centrifugal forces in general relativity. 2. Applications to circular orbits in some familiar stationary axisymmetric space-times,''
  Int.\ J.\ Mod.\ Phys.\ D {\bf 6}, 143 (1997).
  [gr-qc/0106014].



\bibitem{Bini:2002mh} 
  D.~Bini and R.~T.~Jantzen,
  ``Circular holonomy, clock effects and gravitoelectromagnetism: Still going around in circles after all these years,''
  Nuovo Cim.\ B {\bf 117}, 983 (2003)
  [gr-qc/0202085].

\bibitem{Bini:1997ea} 
  D.~Bini, P.~Carini and R.~T.~Jantzen,
   ``The intrinsic derivative and centrifugal forces in general relativity. 1. Theoretical foundations,''
  Int.\ J.\ Mod.\ Phys.\ D {\bf 6}, 1 (1997).
  [gr-qc/0106013].
	

\bibitem{Jantzen:1992rg} 
  R.~T.~Jantzen, P.~Carini and D.~Bini,
  ``The many faces of gravitoelectromagnetism,''
  Ann.\ Phys.\  {\bf 215}, 1 (1992).
  [gr-qc/0106043].

\bibitem{Bini:1994}
  D. Bini, P. Carini, R.T. Jantzen, D. Wilkins, 
  ``Thomas precession in post-Newtonian gravitoelectromagnetism,"
  Phys.\ Rev.\ D {\bf 49}, 2820 (1994).

\bibitem{Thomas:1926dy} 
  L.~H.~Thomas,
  ``The motion of a spinning electron,''
  Nature {\bf 117}, 514 (1926).

\bibitem{Thomas:1927yu} 
  L.~H.~Thomas,
  ``The Kinematics of an electron with an axis,''
  Phil.\ Mag.\  {\bf 3}, 1 (1927).

\bibitem{furry}
W. H. Furry,
``Lorentz Transformation and the Thomas Precession,"
Am. J. Phys. 23, 517 (1955).

\bibitem{shelupsky}
David Shelupsky,
``Derivation of the Thomas Precession Formula,"
Am. J. Phys. 35, 650 (1967).

\bibitem{fischer}
G. P. Fisher,
``The Thomas Precession,"
Am. J. Phys. 40, 1772 (1972).

\bibitem{goedecke}
G. H. Goedecke,
``Geometry of the Thomas precession,"
Am. J. Phys. 46, 1055 (1978).


\bibitem{Ferraro:1999eu} 
  R.~Ferraro and M.~Thibeault,
  ``Generic composition of boosts: an elementary derivation of the Wigner rotation,''
  Eur.\ J.\ Phys.\  {\bf 20}, 143 (1999).

\bibitem{Bakke:2015yia} 
  K.~Bakke, C.~Furtado and A.~M.~de M. Carvalho,
  ``Wigner rotation via Fermi-Walker transport and relativistic EPR correlations in the Schwarzschild spacetime,''
  Int.\ J.\ Quant.\ Inf.\  {\bf 13}, no. 02, 1550020 (2015)
  [arXiv:1504.07432].


\bibitem{O'Donnell:2011am} 
  K.~O'Donnell and M.~Visser,
  ``Elementary analysis of the special relativistic combination of velocities, Wigner rotation, and Thomas precession,''
  Eur.\ J.\ Phys.\  {\bf 32}, 1033 (2011)
  [arXiv:1102.2001 [gr-qc]].

\bibitem{Rhodes:2003id} 
  J.~A.~Rhodes and M.~D.~Semon,
  ``Relativistic velocity space, Wigner rotation and Thomas precession,''
  Am.\ J.\ Phys.\  {\bf 72}, 943 (2004)
  [gr-qc/0501070].

\bibitem{Han:1987gj} 
  D.~Han, Y.~S.~Kim and D.~Son,
  ``Thomas Precession, Wigner Rotations, and Gauge Transformations,''
  Class.\ Quant.\ Grav.\  {\bf 4}, 1777 (1987).

\bibitem{Wigner:1939cj} 
  E.~P.~Wigner,
  ``On Unitary Representations of the Inhomogeneous Lorentz Group,''
  Annals Math.\  {\bf 40}, 149 (1939)
  [Nucl.\ Phys.\ Proc.\ Suppl.\  {\bf 6}, 9 (1989)].


\bibitem{Bini:2016iym} 
  D.~Bini, A.~Geralico and R.~T.~Jantzen,
  ``Gyroscope precession along bound equatorial plane orbits around a Kerr black hole,''
  Phys.\ Rev.\ D {\bf 94},  064066 (2016)
  [arXiv:1607.08427 [gr-qc]].



\bibitem{Gair:2005is} 
  J.~R.~Gair, D.~J.~Kennefick and S.~L.~Larson,
  ``Semi-relativistic approximation to gravitational radiation from encounters with black holes,''
  Phys.\ Rev.\ D {\bf 72}, 084009 (2005)
  Erratum: [Phys.\ Rev.\ D {\bf 74}, 109901 (2006)]
  [gr-qc/0508049].



\bibitem{Damour:2014afa} 
  T.~Damour, F.~Guercilena, I.~Hinder, S.~Hopper, A.~Nagar and L.~Rezzolla,
  ``Strong-Field Scattering of Two Black Holes: Numerics Versus Analytics,''
  Phys.\ Rev.\ D {\bf 89}, 081503 (2014)
  [arXiv:1402.7307 [gr-qc]].


\bibitem{Ruangsri:2015cvg} 
  U.~Ruangsri, S.~J.~Vigeland and S.~A.~Hughes,
  ``Gyroscopes orbiting black holes: A frequency-domain approach to precession and spin-curvature coupling for spinning bodies on generic Kerr orbits,''
  Phys.\ Rev.\ D {\bf 94}, 044008 (2016)
  [arXiv:1512.00376 [gr-qc]].


\bibitem{ferrbini}
F. de Felice and D. Bini,
``Classical Measurements in Curved Space-Times,"
 Cambridge University Press, Cambridge, 2010.


\bibitem{Misner:1974qy} 
  C.~W.~Misner, K.~S.~Thorne and J.~A.~Wheeler,
  ``Gravitation,''
  San Francisco 1973.

\bibitem{Chandrasekhar:1985kt} 
  S.~Chandrasekhar,
   ``The mathematical theory of black holes,''
  Oxford, Clarendon, UK, 1985.


\bibitem{bcjkorea}
D. Bini, P. Carini and R.T. Jantzen,
   ``Applications of Gravitoelectromagnetism to Rotating Spacetimes,"  
   J. Korean Phys. Soc. 25, S190 (1992).


\bibitem{Carter:1968ks} 
  B.~Carter,
  ``Hamilton-Jacobi and Schrodinger separable solutions of Einstein's equations,''
  Commun.\ Math.\ Phys.\  {\bf 10}, 280 (1968).


\bibitem{Glampedakis:2002ya} 
  K.~Glampedakis and D.~Kennefick,
  ``Zoom and whirl: Eccentric equatorial orbits around spinning black holes 
    and their evolution under gravitational radiation reaction,''
  Phys.\ Rev.\ D {\bf 66}, 044002 (2002)
  [gr-qc/0203086].



\bibitem{Levin:2008yp} 
  J.~Levin and G.~Perez-Giz,
  ``Homoclinic Orbits around Spinning Black Holes. I. Exact Solution for the Kerr Separatrix,''
  Phys.\ Rev.\ D {\bf 79}, 124013 (2009)
  [arXiv:0811.3814 [gr-qc]].


\bibitem{Gradshteyn}
I.~S.~Gradshteyn and I.~M.~Ryzhik, 
``Table of Integrals, Series, and Products,'' Ed. by A.~Jeffrey
and D.~Zwillinger, Academic Press, New York, 6th edition, 2000.


\bibitem{marck1}
  J.A. Marck, 
 ``Parallel-tetrad on null geodesics in Kerr-Newman space-time,"
  Phys.\ Lett.\ A {\bf 97}, 140 (1983).

\bibitem{marck2}
N. Kamran and J.A. Marck,
 ``Parallel-propagated frame along the geodesics of the metrics admitting a Killing–Yano tensor,"
 J.\ Math.\ Phys.\ {\bf27}, 1589 (1986).


\bibitem{Bini:2011zzc} 
  D.~Bini and A.~Geralico,
  ``Spin-geodesic deviations in the Kerr spacetime,''
  Phys.\ Rev.\ D {\bf 84}, 104012 (2011)
  [arXiv:1408.4952 [gr-qc]].

\bibitem{Bini:2016xqg} 
  D.~Bini and B.~Mashhoon,
  ``Relativistic gravity gradiometry: the Mashhoon-Theiss effect,''
  arXiv:1607.05473 [gr-qc].

\bibitem{Akcay:2016dku} 
  S.~Akcay, D.~Dempsey and S.~Dolan,
  ``Spin-orbit precession for eccentric black hole binaries at first order in the mass ratio,''
  arXiv:1608.04811 [gr-qc].


\bibitem{Bini1999} 
 D. Bini, R.T. Jantzen and A. Merloni,
  ``Geometric interpretation of the Frenet-Serret frame description of 
    circular orbits in stationary axisymmetric spacetimes,"
    Class.\ Quantum Grav.\ {\bf16}, 1333 (1999).


\bibitem{Bini:2016ubc} 
  D.~Bini and A.~Geralico,
  ``Scattering by a Schwarzschild black hole of particles undergoing drag force effects,''
  Gen.\ Rel.\ Grav.\  {\bf 48}, 101 (2016).


\end{thebibliography}
\end{document}